\documentstyle[preprint,aps,epsf]{revtex}
\def\be {\begin{eqnarray}}
\def\ee {\end{eqnarray}}
\def\gord{$ \raisebox{-.3ex}{$\stackrel{>}{_{\sim}}$} $}
\def\beq {\begin{equation}}
\def\eeq {\end{equation}}
\def\del {\partial}
\def\bi {\begin{itemize}}
\def\ei {\end{itemize}}
\def\ben {\begin{enumerate}}
\def\een {\end{enumerate}}
\def\ni {\noindent}
\def\gam2{\gamma_{\frac{1}{2}}}
\newcommand{\btem}{\bibitem}

\newcommand{\vm}[1]{\mbox{\bf#1}}
\newcommand{\vms}[1]{\mbox{\scriptsize{\bf#1}}}
\def\thalf{{\textstyle{\frac{1}{2}}}}
\def\tquar{{\textstyle{\frac{1}{4}}}}
\def\thquar{{\textstyle{\frac{3}{4}}}}
\def\teighth{{\textstyle{\frac{1}{8}}}}
\def\thfive{{\textstyle{\frac{3}{5}}}}

\def\fpj{\hspace{-.7cm}}
\def\vol {{\cal V}}
\def\tr  {{\rm Tr}}
\def\ni {\noindent}

\def\gd {\frac{\gamma'}{\phi}}
\def\gdd {\gamma''}

\def\gtdd {\gamma_\frac12''}
\begin{document}
\preprint{NUC-MINN-96/9-T,DOE/ER/40427-14-N96}
\title{Kaon Zero-Point Fluctuations in Neutron Star Matter}
\author{Vesteinn Thorsson}
\address{Department of Physics, University of Washington \\
Box 351560, Seattle, Washington 98195-1560, USA\cite{byline} \\
and\\
NORDITA, Blegdamsvej 17, DK-2100 Copenhagen {\O}, Denmark\\
}
\author{Paul J. Ellis }
\address{ School of Physics and Astronomy, University of Minnesota \\
Minneapolis, MN 55455-0112, USA}
\date{\today}
\maketitle
\begin{abstract}

We investigate the contribution of zero-point motion, 
arising from fluctuations in kaon modes, to the ground
state properties of neutron star matter containing
a Bose condensate of kaons.  
The zero-point energy is derived via the thermodynamic partition
function, by integrating out fluctuations for an arbitrary value of 
the condensate field.  It is shown that the vacuum counterterms of the chiral 
Lagrangian ensure the cancellation of divergences dependent
on $\mu$, the charge chemical potential, which may be regarded
as an external vector potential.  The total grand potential,
consisting of the tree-level potential, the zero-point contribution,
and the counterterm potential, is extremized
to yield a locally charge neutral, beta-equilibrated and
minimum energy ground state. In some regions of parameter space we
encounter the well-known problem of a complex effective potential.
Where the potential is real and solutions can be obtained, the contributions 
from fluctuations are found to be small in comparison
with tree-level contributions.    

\end{abstract}
\pacs{PACS numbers: 26.60.+c, 12.39.Fe, 21.65.+f, 97.60.Jd}

%
%
\section{Introduction}

In recent years the idea that the ground state of nuclear matter at high 
density contains a Bose condensate of kaons has received considerable 
attention.  The existence of this novel state of matter is suggested 
by the attractive {\it s}-wave kaon-nucleon interactions found in low energy
effective chiral Lagrangians, which, by construction, preserve the
symmetries of QCD. This is supported experimentally by kaonic atom 
data which indicate \cite{bat} a strongly attractive $K^-$ optical 
potential. Attractive kaon-nucleon interactions imply that 
in dense matter, such as that in the interior of neutron stars, 
the effective kaon mass is lower than in free space.
It could thus be energetically favorable for leptons in the 
matter to convert to kaons via strangeness changing interactions, in which 
case the lowest energy kaon mode becomes macroscopically populated,
forming a condensate.  

Since the original suggestion by Kaplan and Nelson\cite{kapnel},
studies of kaon condensation and kaon propagation in the nuclear 
medium can broadly be divided into three
categories:  studies of the basic interactions,
consequences for heavy-ion collisions and astrophysical implications.

A number of studies have been aimed at improving 
the description of kaon-nucleon interactions over 
that of the original model. 
Brown, Lee, Rho and Thorsson\cite{blrt} included
all terms to next-to-leading order (${\cal O}(q^2)$) 
in chiral perturbation theory,   
and fitted the expansion parameters to $KN$ scattering lengths.
Lee, Brown, Min and Rho\cite{lbmr} included all terms to 
next-to-next-to-leading order (${\cal O}(q^3)$) 
in chiral perturbation theory and also fitted kaonic atom data.
Lee {\it et al.} have also included the $\Lambda(1405)$ (which
dominates the low-energy $K^-p$ interaction), 
four-baryon 
interactions, and Pauli blocking. 
Other noteworthy studies are those of Ref.\cite{ppt},
concerning the effect of kaon-nucleon correlations, 
and the work of the Kyoto group ( see e.g. Ref.\cite{kyoto} ).
Waas, Kaiser, and Weise\cite{waas}
have recently investigated kaon modes in matter
using a multichannel K-matrix analysis. 
For our purposes, it is worth mentioning 
that the majority of these more refined investigations have 
shown that the basic mechanism for kaon condensation, namely 
the existence of an attractive interaction, is robust.

As regards heavy ion collisions, strangeness is conserved on the 
relevant timescales and a kaon condensate
is not likely to form.  However, modification of 
the kaon modes following from the chiral Lagrangian
could have consequences in kaon flow\cite{likoli},
for subthreshold production of kaons\cite{subt},
or for the $\phi$-meson peak in dilepton spectra\cite{phi}.

Another major direction that investigations have taken
is to study the astrophysical consequences of the 
formation of a kaon condensate (the effects appear to be reduced if
hyperons are also present \cite{hyp}, a possibility that we do not 
consider here). The presence of such a condensate
considerably enhances the proton concentration which leads to
a softer equation of state (EOS) and a lower maximum mass for the 
cold neutron 
star. The actual value of this maximum mass is dependent on the nuclear
EOS, but for the non-relativistic EOS employed here
it is quite close \cite{blrt,bkrt,tpl} to the 
lower limit of $1.44M_\odot$ set by PSR 1913+16 \cite{Taylor}.
An interesting effect occurs when the neutron star has just formed
since electron neutrinos are trapped for  about 10--15 s (also the 
entropy/baryon is 
$\sim1-2$, but this plays a lesser role). The changes in the stellar 
composition induced by trapping increase the threshold for kaon 
condensation. This stiffens the EOS and allows a maximum mass 
which is $\sim0.1-0.2M_\odot$ larger than in the absence of trapped 
neutrinos. Thus there is the possibility of metastable neutron stars
which become black holes after about 10--15 s when the trapped neutrinos have 
left \cite{tpl,keiljan,pcl,subside,newborn}. This could account for the 
fact that fewer pulsars are observed in supernova remnants than
expected from conventional scenarios \cite{bb}. As regards dynamical 
simulations \cite{keiljan,burr} of the delayed collapse of hot 
neutron stars into black holes, Baumgarte, Shapiro and Teukolsky\cite{bst} 
have recently examined the effect of kaon condensates.
Here the cold EOS is supplemented
with simple assumptions about corrections for finite temperature.

One-loop calculations to date are
valid only up to the density at which the condensate
forms where the condensate amplitude is vanishingly small. 
In order to go beyond threshold we
analyze the fluctuations around the kaon condensate to one loop order and
include the full non-linearities of the Lagrangian. This is of
general theoretical interest, as well as in the particular context of 
neutron stars. While we shall restrict our numerical analysis here to the 
zero-point contribution at zero temperature, the formalism also provides
a consistent treatment of the thermal contributions due to kaons.
These would be needed in studies of newborn neutron stars.

In the work which follows, we shall make the following 
approximations.   Nucleons are treated non-relativistically, whereas
a more complete treatment would involve a relativistic treatment to one 
loop order. In the meson sector, we restrict the
discussion to kaonic degrees of freedom.  
We thus omit from the discussion fluctuations in the
pion and eta sectors.  
Physically one might argue that ignoring
non-kaonic pseudoscalar degrees of freedom is justified since one expects
that the corresponding modes are less
modified in nuclear matter than those of kaons.
The methods developed here
could be used to include the full octet of pseudoscalar 
fluctuations, but this would further complicate the present analysis.
Lastly, we have sought to treat the chiral Lagrangian  
at the most simple level starting from the Lagrangian
of Kaplan and Nelson \cite{kapnel}, although one would wish to 
include the additional terms of Refs. \cite{blrt,lbmr} at a later stage.

The paper is organized as follows.  In Sec. II,
we present the basic formalism, including the evaluation of 
the kaon partition function, a discussion of divergences
in the zero-point energy, the nucleon Hamiltonian, and the
equilibrium conditions.  In Sec. III we compare numerical results 
in neutron star matter with and without zero-point fluctuations. 
We discuss the critical density for kaon 
condensation, the composition of the star and the EOS.
Following the Conclusions in Sec. IV are three Appendices.
Appendix A concerns the cancellation of divergences dependent
on the chemical potential in a toy model, while Appendix B
discusses the case of the actual chiral Lagrangian. 
In Appendix C we derive the critical density from the 
kaon propagator to one-loop order, and show that 
it is the same as that obtained by expanding the zero-point
energy in terms of the kaon fields.

%
%
%
%

\section{Formalism}

Our principal interest is the evaluation of the impact of the kaon 
zero-point energy contribution on the structure of neutron stars. In 
Subsec. A we obtain the partition function for the chiral Lagrangian of 
Kaplan and Nelson \cite{kapnel} to one loop order and in Subsec. B we 
discuss the elimination of the divergences.
For the nucleons alone we use
a simple non-relativistic Hamiltonian which should be sufficient for 
present purposes, as outlined in Subsec. C. The stellar equilibrium 
conditions 
that need to be satisfied are given in Subsec. D.

\subsection{Kaon Partition Function}

The $SU(3)\times SU(3)$ chiral Lagrangian involving the 
pseudoscalar meson octet and baryon octet is\cite{kapnel,blrt,lbmr,jm}  
\be
{\cal L}_K &=& {\cal L}^{(1)}+{\cal L}^{(2)}+{\cal L}^{(0)}_{ct}
+{\cal L}^{(3)}_{ct}+{\cal L}^{(4)}_{ct}
\nonumber \\
{\cal L}^{(1)} &=&   i {\rm Tr} \bar{B} v \cdot \del B +
i{\rm Tr}\bar{B}v^{\nu}[V_{\nu},B] + 
2 D {\rm Tr} \bar{B} S^\nu \{ A_\nu, B \} +
2 F {\rm Tr} \bar{B} S^\nu  [ A_\nu, B ] +
{\cal L}^{(1)}_{ct} 
\nonumber \\
{\cal L}^{(2)} &=& \tquar f^2 {\rm Tr}\partial_{\nu}U\partial^{\nu}U\!+
C{\rm Tr}\,m_q(U+U^{\dagger}\!-2)  +a_1{\rm Tr}\bar{B}(\xi m_q\xi+\!h.c.)B
 \nonumber\\
&&\quad +a_2{\rm Tr}\bar{B}B(\xi m_q\xi+h.c.)
+a_3\left\{{\rm Tr}\bar{B}B\right
\}
\left\{{\rm Tr}(m_qU+h.c.)\right\} + {\cal L}^{(c,d,f)} +
{\cal L}^{(2)}_{ct}\;.
\label{pje0}
\ee
The bracketed superscripts denote the order in the low 
momentum expansion defining the effective field theory.
The various one-loop counterterms, with subscripts {\it ct}, 
are discussed in Appendices A and B; here in the main text we 
will simply put them all together in a single ${\cal L}_{ct}$.
The additional terms denoted by ${\cal L}^{(c,d,f)}$ are given in 
Refs. \cite{blrt,lbmr} and those with coefficients $d$ can contribute 
to $s$-wave kaon-nucleon interactions. In the interests of 
simplicity we set  
${\cal L}^{(c,d,f)}=0$ and so use as a starting point the 
Lagrangian of Kaplan and Nelson\cite{kapnel}.
The Lagrangian (\ref{pje0}) is derived in the 
limit of heavy baryons, where $v^\nu=(1,{\bf 0})$ is 
the velocity four-vector of the baryons,
and $S^\nu$ is the spin operator ($v \cdot S=0$,$S^2=-\frac34$). 

In Eq. (\ref{pje0}) , $U=\xi^2$ is the non-linear field involving the 
pseudoscalar meson octet, 
from which we retain only the $K^{\pm}$ contributions--
\begin{equation}
U=\exp\left(\frac{\sqrt{2}i}{f}M\right)\quad;\quad
M=\left(\matrix{0&0&K^+\cr
0&0&0\cr K^-&0&0\cr}\right)\;,
\end{equation}
and from the baryon octet we retain just the nucleon contributions
\begin{equation}
B=\left(\matrix{0&0&p\cr
0&0&n\cr
0&0&0\cr}\right)\;.
\end{equation}
Consequently the $p$-wave interactions in ${\cal L}^{(1)}$ with coefficients
$D$ and $F$ do not enter 
($A_\nu=\frac{i}{2}\{\xi^\dagger,\del_\nu \xi\}$). 
In Eq. (\ref{pje0}), the quark mass matrix
$m_q={\rm diag}(0,0,m_s)$;  {\it i.e.}, only the mass of the strange quark 
is taken to be non-zero. For the mesonic vector current, $V_{\nu}$,
only the time component  survives in an infinite system with
$V_0=\frac12 [\xi^\dagger,\del_0 \xi]$. 
Also, the pion decay constant $f=93$ MeV, and $C, a_1, a_2$, and $a_3$ 
are constants.
After some algebra, the relevant part of ${\cal L}_K$ can be written
\be
{\cal L}_K&=&\thalf(1+\gamma)\del_{\nu}K^+\del^{\nu}K^-
+\frac{(1-\gamma)}{2f^2\Theta^2}\left[(K^+\del_{\nu}K^-)^2
+(K^-\del_{\nu}K^+)^2\right]\nonumber\\
&&\quad+\gam2
\left[ibK^+\stackrel{\leftrightarrow}{\del_0}K^-
-(m_K^2+c)K^+K^-\right]+{\cal L}_{ct}\;. \label{lkaon}
\ee
Here, the kaon mass is given by $m_K^2=2Cm_s/f^2$, and we have 
employed the following definitions:
\be
&&\fpj \Theta^2=\frac{2}{f^2}K^+K^-\quad;\quad
\gamma=\left(\frac{\sin\Theta}{\Theta}\right)^{\!\!2}\quad;\quad
\gam2=\left(\frac{\sin\thalf\Theta}{\thalf\Theta}\right)^{\!\!2}\nonumber\\
&&\fpj b=(1+x)\frac{n}{4f^2}\quad;\quad c=(a_1x+a_2+2a_3)\frac{m_sn}
{f^2}\;.
\label{defs}
\ee
Expectation values of nucleon operators are evaluated
in the mean-field approximation. 
The total nucleon number density is $n=n_p+n_n$ and the proton 
fraction is $x=n_p/n$. Note that since we shall use a non-relativistic 
model for the nucleons we have not distinguished here between the 
scalar density and the number density in the expression for $c$.  
The nucleon kinetic energy in ${\cal L}^{(1)}$ will 
be considered in Subsection C and so is not included in Eq. (\ref{lkaon}).

We have not included in Eq. (\ref{lkaon}) terms which simply give a constant
shift to the baryon masses; they indicate that $a_1m_s=-67$ MeV and
$a_2m_s=134$ MeV, using the hyperon-nucleon mass differences. The remaining
constant $a_3m_s$ is not accurately known, and we shall use values in the
range $-134$ to $-310$ MeV corresponding to 0 to 20\% strangeness content
for the proton. The corresponding range for the kaon-nucleon sigma term,
\begin{equation}
\Sigma^{KN}=-\thalf(a_1+2a_2+4a_3)m_s \,,
\end{equation}
is 167--520 MeV. Some 
guidance is provided by recent lattice gauge 
simulations~\cite{dong}, which find that the strange quark condensate in the 
nucleon is large,
{\it i.e.}, $\langle N|\overline ss|N\rangle = 1.16\pm0.54$. 
{}From the relation
$m_s\langle\overline ss\rangle_p = - 2(a_2+a_3)m_s$ and using $m_s=150$ MeV,
we obtain $a_3m_s = -(220 \pm40)$ MeV, which is in the middle of our range
of values.

In order to determine the kaon partition function  
we generalize the finite temperature procedure outlined by
Kapusta~\cite{kapusta} and Benson, Bernstein and Dodelson \cite{bbd}. First, by
studying  the invariance of the Lagrangian under the transformation 
$K^{\pm}\rightarrow K^{\pm}e^{\pm i\alpha}$, the conserved current density 
can be identified. The zero component, {\it i.e.} the charge density, is
\begin{equation}
j_0=i\gamma K^+\stackrel{\leftrightarrow}{\del_0}K^- + 
2\gam2 bK^+K^-\;.
\end{equation}
Next, we transform to real fields $\phi_1$ and $\phi_2$,
\begin{equation}
K^{\pm}=(\phi_1\pm i\phi_2)/\sqrt{2}\;.
\label{kphi}
\end{equation}
The Lagrangian becomes
\be
{\cal L}_K&=&\thalf A\del_{\nu}\phi_1\del^{\nu}\phi_1
+\thalf B\del_{\nu}\phi_2\del^{\nu}\phi_2+C\del_{\nu}\phi_1\del^{\nu}\phi_2
\nonumber\\
&&\quad+\gam2\left[b\phi_1\stackrel{\leftrightarrow}{\del_0}\phi_2
-\thalf(m_K^2+c)(\phi_1^2+\phi_2^2)\right]+{\cal L}_{ct}\;,
\label{lphi}
\ee
where
\beq
A=\frac{\phi_1^2+\gamma\phi_2^2}{\phi_1^2+\phi_2^2}\quad;\quad
B=\frac{\phi_2^2+\gamma\phi_1^2}{\phi_1^2+\phi_2^2}\quad;\quad
C=\frac{(1-\gamma)\phi_1\phi_2}{\phi_1^2+\phi_2^2}\;. \label{abc}
\eeq
Since, to our order of approximation, the counterterm Lagrangian, 
${\cal L}_{ct}$, is to be evaluated with constant mean fields it
plays no role in determining the conjugate momenta. These are 
\be
\pi_1&=&A\partial_0\phi_1+C\partial_0\phi_2-\gam2 b\phi_2\nonumber\\
\pi_2&=&B\partial_0\phi_2+C\partial_0\phi_1+\gam2 b\phi_1\;.
\ee
The Hamiltonian density is
${\cal H}_K=\pi_1\partial_0\phi_1+\pi_2\partial_0\phi_2
-{\cal L}_K$, and the partition function of the grand canonical ensemble can 
then be written as the functional integral
\begin{eqnarray}
Z_K&=&\int[d\pi_1][d\pi_2]\int_{periodic}[d\phi_1][d\phi_2]\nonumber\\
&&\quad\times\exp\left\{\int\limits_0^{\beta}d\tau\int d^3x\left(
i\pi_1\frac{\partial\phi_1}{\partial\tau}+
i\pi_2\frac{\partial\phi_2}{\partial\tau}-{\cal H}_K(\phi_i,\pi_i)
+\mu j_0(\phi_i,\pi_i)\right)\right\}\;.\label{zint}
\end{eqnarray}
Here $\mu$ is the chemical potential associated with the conserved charge 
density, $\beta=T^{-1}$ is the inverse temperature and the fields obey 
periodic boundary conditions in the imaginary time
$\tau=it$, namely $\phi_i(\vm{x},0)=\phi_i(\vm{x},\beta)$.

The Gaussian integral over momenta in Eq. (\ref{zint}) can be performed
yielding
\be
Z_K&=&N\int[d\phi_1][d\phi_2]\gamma^{\frac{1}{2}}e^{-S}
\qquad{\rm with}\nonumber\\
S&=&\thalf\int\limits_0^{\beta}d\tau\int d^3x\Biggl\{
A\left(\frac{\del\phi_1}{\del\tau}\right)^{\!\!2}
+B\left(\frac{\del\phi_2}{\del\tau}\right)^{\!\!2}
+2C\frac{\del\phi_1}{\del\tau}\frac{\del\phi_2}{\del\tau}\nonumber\\
&&+A(\bbox{\nabla}\phi_1)^2+B(\bbox{\nabla}\phi_2)^2
\!+2C\bbox{\nabla}\phi_1\cdot\bbox{\nabla}\phi_2
+2i(\gamma\mu+\gam2 b)\phi_2\stackrel{\leftrightarrow}{\del_{\tau}}\phi_1
\nonumber\\
&&+\left[\gam2(m_K^2+c-2\mu 
b)-\gamma\mu^2\right](\phi_1^2+\phi_2^2)-2{\cal L}_{ct}\Biggr\}\;. 
\label{action}
\ee
The dependence
of the action (\ref{action}) upon the chemical potential $\mu$ can also be 
obtained by gauging the fields as discussed in Appendices A and B.

Next the fields are decomposed into constant condensate parts and fluctuations
according to
\begin{eqnarray}
\phi_1&=&\bar{\phi}_1+\phi_1'=f\theta\cos\alpha
+\sqrt{\frac{\beta}{\vol}}\sum_{n,\vms{p}}
e^{i(\vms{p}\cdot\vms{x}+\omega_n\tau)}\phi_{1,n}(\vm{p})\nonumber\\
\phi_2&=&\bar{\phi}_2+\phi_2'=f\theta\sin\alpha
+\sqrt{\frac{\beta}{\vol}}\sum_{n,\vms{p}}
e^{i(\vms{p}\cdot\vms{x}+\omega_n\tau)}\phi_{2,n}(\vm{p})\;.
\label{decomp}
\end{eqnarray}
Here $\vol$ denotes the volume of the system. 
In the Fourier expansion of the fluctuating part
$\phi_{1,0}(\vm{p}=0)=\phi_{2,0}(\vm{p}=0)=0$ and the Matsubara frequency
$\omega_n=2\pi nT$. Since we wish to go to one loop order, ${\cal L}_{ct}$
is evaluated with constant condensate fields and we write the
contribution as a potential $V_{ct}$. The remainder of $S$ must be expanded 
to second order in the fluctuations.  The partition function 
can then be written in the form
\beq
Z_K=Ne^{-\beta V(V_{tree}+V_{ct})}\int \prod_{n,\vms{p}}
[d\phi_{1,n}(\vm{p})] [d\phi_{2,n}(\vm{p})]
\gamma^{\frac{1}{2}}e^{-S_{loop}}\;. \label{zone}
\eeq
Introducing the definitions
\be
&&\epsilon(\theta) = 2( m_K^2 + c - 2 \mu b ) f^2 \sin^2\!\thalf\theta
- \thalf\mu^2 f^2 \sin^2\theta  \nonumber\\
&&\rho(\theta) = -\frac{\del\epsilon}{\del\mu} = 4 b f^2 \sin^2\!\thalf\theta
+ \mu f^2 \sin^2 \theta \;,
\ee
the tree potential is
\beq
V_{tree}=\epsilon(\theta)\;.
\label{vtree}
\eeq
The one loop contribution to the action then becomes
\be
S_{loop}&=&\thalf\sum_{n,\vms{p}}\Bigl(\phi_{1,-n}(-\vm{p}),
\phi_{2,-n}(-\vm{p})\Bigr)
\bbox{D}\left(\matrix{\phi_{1,n}(\vm{p})\cr\phi_{2,n}(\vm{p})\cr}\right)\;,
\quad{\rm where}\nonumber\\
\beta^{-2}f^2 D_{11}  &=& Af^2( \omega_n^2 + {\bf p}^2 )
- \omega_n\theta 
\left( \frac{\rho}{\theta^2}\right)'\sin2\alpha+
\epsilon'' \cos^2\!\alpha+\epsilon' \frac{\sin^2\!\alpha}{\theta}\nonumber \\
\beta^{-2}f^2 D_{22}  &=& Bf^2( \omega_n^2 + {\bf p}^2 )
+ \omega_n\theta 
\left( \frac{\rho}{\theta^2}\right)'\sin2\alpha+
\epsilon''\sin^2\!\alpha + \epsilon'\frac{\cos^2\!\alpha}{\theta}\nonumber \\
\beta^{-2}f^2 D_{12}  &=& Cf^2( \omega_n^2 + {\bf p}^2 )
+2 \omega_n \left( \frac{\rho}{\theta^2} + \theta\cos^2\!\alpha
\left( \frac{\rho}{\theta^2}\right)'  \:\right)
+ \left(\epsilon'' - \frac{1}{\theta}\epsilon'\right)\sin\alpha\cos\alpha
\nonumber\\
\beta^{-2}f^2 D_{21}  &=& Cf^2( \omega_n^2 + {\bf p}^2 )
-2 \omega_n \left( \frac{\rho}{\theta^2} + \theta\sin^2\!\alpha
\left( \frac{\rho}{\theta^2}\right)'  \:\right)
+ \left(\epsilon'' - \frac{1}{\theta}\epsilon'\right)\sin\alpha\cos\alpha \;,
\label{Dmatrix}
\ee
with a prime denoting a partial derivative with respect to $\theta$.
Here the quantities $A,\ B$ and $C$ of Eq. (\ref{abc}) are to be evaluated 
with the condensate fields, giving
\beq
A=1+(\gamma-1)\sin^2\!\alpha\quad;\quad
B=1+(\gamma-1)\cos^2\!\alpha\quad;\quad
C=(1-\gamma)\sin\alpha\cos\alpha\;,
\eeq
and $\gamma=(\sin\theta/\theta)^2$. The latter replacement should also be 
made in the measure of the integral (\ref{zone}), to the order considered.
Then the integration can be carried out and the grand potential $\Omega_K$ is
given by
\beq
\frac{\Omega_K}{\vol}=-\frac{1}{\beta \vol}\ln Z_K=V_{tree}+V_{ct}
+\frac{1}{2\beta \vol}\sum_{n,\vms{p}}\ln\frac{1}{\gamma}{\rm det}\bbox{D}\;. 
\label{grand1}
\eeq
Evaluating the determinant yields
\be
\frac{1}{\gamma}{\rm det}\bbox{D}&=& \beta^4\left[(\omega_n^2 + \vm{p}^2 )^2
+ 2 \omega_n^2 ( c_1 + c_2 + c_3 ) + 2 \vm{p}^2 ( c_2 + c_3 )
+ 4 c_2 c_3 \right] \nonumber\\
&\equiv& \beta^2( \omega_n^2 + E_+^2 ) \cdot \beta^2(  \omega_n^2 + E_-^2 )\;,
\label{detD}
\ee
where
\be
&&c_1 = \frac{2\rho}{f^4 \gamma \theta^2 }
\frac{\del}{\del \theta} \left( \frac{\rho}{\theta} \right)  \quad,\quad
c_2 = \frac{1}{2f^2} \frac{\del^2 \epsilon}{\del \theta^2}   \quad,\quad
c_3 = \frac{1}{2f^2\gamma\theta} \frac{\del\epsilon}{\del\theta} 
\label{cs} \\
&&E_{\pm}^2(\vm{p}) =
{\vm p}^2 + c_1 + c_2 + c_3 \pm
\sqrt{  c_1 ( 2 {\vm p}^2 + c_1 + 2 c_2 + 2 c_3 ) + ( c_2 - c_3 )^2 }\;.
\label{disp}
\ee
Notice that there is no longer any dependence on the condensate phase 
$\alpha$. The sum over the Matsubara frequencies in Eq. (\ref{grand1}) 
can then be evaluated in the usual way \cite{kapusta} to yield
\be
&&\Omega_K {\vol}^{-1}=V_{tree}+V_{ct}+V_{zp}+V_{thermal}\quad{\rm with}
\nonumber\\
&&V_{zp} = \int \frac{ d^3p}{(2\pi)^3} \:\thalf
[E_+(\vm{p}) + E_-(\vm{p})]
\label{vzp}
\\
&&V_{thermal} = \beta^{-1} \int \frac{ d^3p}{(2\pi)^3}
\ln( 1 - e^{-\beta E_+(\vms{p})})(1 - e^{-\beta E_-(\vms{p})} ) \;.
\label{vthermal}
\ee
We are interested in zero temperature here, so $V_{thermal}$ may be dropped. 

The zero-point energy $V_{zp}$, or equivalently, the 
one-kaon-loop contribution to the effective potential\cite{pokorski}, 
can be represented as an
infinite series of diagrams, some of which are shown in Fig. 1.
Vertices are given by the Lagrangian (\ref{lkaon}).
External legs represent nucleon or kaon mean fields.

It is interesting to note the chiral order of the diagrams 
contributing to the effective potential.
Consider the {\it scattering 
amplitudes} corresponding to the diagrams of Fig. 1, where
the external lines are on their respective  vacuum mass-shells.
The chiral order, $\nu$, of a given scattering amplitude is 
given by the Weinberg\cite{wein} counting rule,
according to which an amplitude involving $E_N$ 
external nucleon lines and $E_K$ external kaon lines
is proportional to $Q^\nu$, where
$Q$ is a characteristic small momentum
scale involved in the process and
\be
\nu=2+2L-\thalf E_N +\sum_i \left(d_i +\thalf n_i -2\right)\;,
\label{count}
\ee
where $L$ is the number of loops, the sum is over all
vertices $i$, $d_i$ the number of derivatives that act
on the $i$th vertex and $n_i$ the number of nucleon lines attached to
the $i$th vertex.
In this case Fig. 1a gives $\nu=4$.
Furthermore, it easy to convince oneself that adding a vertex
from ${\cal L}^{(1)}$, corresponding to the $b$ term in
Eq. (\ref{lkaon}), will decrease $\nu$ by 1.
On the other hand, adding a vertex from  ${\cal L}^{(2)}$,
corresponding to the kinetic, mass, or $c$ terms in 
Eq. (\ref{lkaon}) leaves $\nu$ invariant.
The fact that these  scattering amplitudes give $-\infty<\nu \leq 4$
due to the ${\cal L}^{(1)}$ interaction might appear to be
grounds for concern.
However, exactly the same apparent counting problem arises when one
considers the computation of the tree-level in-medium
kaon propagator.
The correct in-medium propagator is indeed obtained by
summing an infinite series\cite{tw} in diagrams that here
correspond to amplitudes with decreasing powers of $\nu$.
The evaluation of the effective potential is entirely analogous.
We conclude that while Eq. (\ref{count}) provides an effective 
way to organize terms in order of importance for a
given scattering process, it cannot be naively 
applied in the same way to the effective potential in the
case at hand.

%
%

\subsection{Zero-point Divergences}

As shown in Appendices A and B, $V_{ct}$ 
cancels the ultraviolet divergences 
in $V_{zp}$. 
For our purposes the zero-point energy is most conveniently written
\beq
V_{zp} = \sqrt{\frac{1}{2}}\int \frac{ d^3p}{(2\pi)^3} \:
\left( \vm{p}^2 + c_1 + c_2 + c_3 +
\sqrt{(\vm{p}^2 + 2 c_2)(\vm{p}^2 + 2 c_3)}     
\right)^\frac12\;.
\label{zpdiv}
\eeq
We regularize the infinities that occur here by introducing an ultraviolet 
cutoff $\Lambda$ and requiring $|\vm{p}|\leq\Lambda$.
(Dimensional regularization is discussed in Appendix B.)
The counterterm potential $V_{ct}$ following from ${\cal L}_{ct}$ 
(see Appendix B) is
\beq
V_{ct} = - (d_4+f_4) - \thalf ( c_1 + 2 c_2 + 2 c_3 ) (d_2+f_2)
+\thalf [ ( c_2 - c_3 )^2 + \tquar ( c_1 + 2c_2 + 2c_3 )^2 ] (d_0+f_0)\;.
\label{ct}
\eeq
Here we have introduced finite terms $f_0$, $f_2$ and $f_4$ as well as
the quartically, quadratically and logarithmically 
divergent integrals 
\be
d_4 &=& \int_{k \leq \Lambda } \frac{d^3k}{(2\pi)^3} k
= \frac{\Lambda^4}{8 \pi^2 } 
\label{quartdiv} \\
d_2 &=& \int_{k \leq \Lambda } \frac{d^3k}{(2\pi)^3} \frac{1}{2k}
= \frac{\Lambda^2}{8 \pi^2 }
\label{quaddiv} \\
d_0 &=& \int_{k \leq \Lambda } \frac{d^3k}{(2\pi)^3} \frac{1}{4k^3}= 
\frac{1}{8 \pi^2 }\ln\left(\frac{\Lambda}{\kappa}\right)\;.
\label{logdiv}
\ee
Here $\kappa$ is an arbitrary scale introduced so that the argument of the 
logarithm is dimensionless. Note that the quadratically and logarithmically 
divergent pieces of $V_{zp}$ in (\ref{zpdiv}) and 
$V_{ct}$ in (\ref{ct}) are individually dependent upon the chemical 
potential $\mu$.
This dependency on the chemical potential $\mu$ arises
from the functions of the kaon fields multiplying
the kinetic terms in Eq. (\ref{lkaon}).
However, $\mu$-dependent divergences in the terms $V_{zp}$
and $V_{ct}$ cancel, as do the $\mu$-independent divergences
of the vacuum theory.
The cancellation of $\mu$-dependent divergences follows from 
a careful treatment of the counterterm lagrangian which regulates
the vacuum theory.  Further discussion of this point is 
given in Appendices A and B.
Here we note that $\mu$ may be considered a constant 
external vector potential
$A_\nu$, with vanishing spatial components.
The general statement being made is that all Green functions
with constant external $A_\nu$ fields are finite.
It is easy to check this for simple examples related to our
theory, such as scalar electrodynamics. 
The fact that it also holds for a complicated nonlinear theory
expanded about an arbitrary point $\phi_1$, $\phi_2$ in function space
 is almost certainly a consequence of 
the Ward-Takahashi identities associated with the U(1) symmetry
transformation preceding Eq. (7).

In order to determine the finite contributions $f_i$ we apply three 
conditions in the vacuum. First  $\Omega_K$, or equivalently the 
pressure, should be zero in the vacuum and, since there is no condensate,
the loop contribution must vanish.
Secondly, we require the zero-momentum propagator to be of the usual form 
$(\omega^2-m_K^2)^{-1}$, see Appendix C. Thirdly we require that the 
four-point vertex $(K^+K^-)^2/(6f^2)$ of Eq. (C3) be unmodified in vacuum.
These conditions imply
\be
&&V_{zp}+V_{ct}\Big|_{vacuum}=0\nonumber\\
&&\frac{\partial^2}{\partial\theta^2}(V_{zp}+V_{ct})\bigg|_{vacuum}=0
\nonumber\\
&&\frac{\partial^4}{\partial\theta^4}(V_{zp}+V_{ct})\bigg|_{vacuum}=0\;.
\ee
Expanding to fourth order in $\theta$, one finds
\be
&&f_4=\frac{m_K^4}{64\pi^2}\quad;\quad f_2=-\frac{m_K^2}{16\pi^2}\nonumber\\
&&f_0=\frac{1}{8\pi^2}\left(\ln\frac{\kappa}{m_K}+\zeta-\thquar\right)\;,
\ee
where using our three momentum cutoff, the constant $\zeta=\ln2-1/4=0.443$; 
interestingly a closely similar value for the constant is obtained with
dimensional regularization, namely $\zeta=\thalf\Psi(3)=0.461$ (Here 
$\Psi$ represents the digamma function \cite{as}). Notice that the arbitary
scale $\kappa$ only occurs in the combination $(d_0+f_0)$ where it cancels 
out. The  disappearance of the arbitary scale is one justification for 
introducing the finite counterterms, we discuss another below.

In evaluating the grand potential 
for the general case numerical integration of Eq. (\ref{zpdiv}) for $V_{zp}$
is required. However below threshold when $\theta=0$ (or just at threshold 
when $\theta$ is vanishingly small) the expressions simplify greatly and
an analytic form can be obtained. 
Explicitly we find $c_1=2(b+\mu)^2$, 
$c_2=c_3=\frac12 (m_K^2+c-2 \mu b-\mu^2)$,
and eigenvalues
\beq
E_{\pm} = \sqrt{{\vm p}^2 + \xi^2} \pm (b+\mu) \quad{\rm with}\quad
\xi^2=b^2 + m_K^2 + c\qquad{\rm for}\ \theta=0\;.
\label{epm}
\eeq
Notice that $(E_{\pm}\mp\mu)$ is simply the tree level energy
of a $K^{\pm}$ meson. In this limit the one loop energy becomes
\be
V_{zp} +V_{ct} &=& \int \frac{ d^3p}{(2\pi)^3}\sqrt{{\vm p}^2+\xi^2}
- (d_4+f_4) - \xi^2(d_2+f_2) 
+ \thalf \xi^4 (d_0+f_0) \nonumber\\
&=& - \frac{1}{64\pi^2} \left[ (m_K^2-3\xi^2) (m_K^2-\xi^2) 
+2\xi^4 \ln \frac{m_K^2}{\xi^2} \right]  
\qquad{\rm for}\ \theta=0\;,
\label{vzp0}
\ee
which vanishes in the vacuum since $\xi=m_K$ there.

%
%

\subsection{Nucleon Hamiltonian}

To describe the nuclear kinetic energy and interactions we employ 
the  non-relativistic approach suggested by 
Prakash, Ainsworth and Lattimer \cite{pal}, which has been rather widely 
employed, {\it e.g.} Refs. \cite{lbmr,tpl}.
Here the Hamiltonian density is written
\beq
{\cal H}_{N}= \thfive E_F^0 u^{\frac{5}{3}}n_0 + V(u) + n(1-2x)^2S(u)\;,
\label{hnuc}
\eeq
where nuclear saturation density is denoted by $n_0$ 
($n_0=0.16 \,{\rm fm}^{-3}$)
and the corresponding 
Fermi momentum and energy are $p_F^0=(3\pi^2n_0/2)^{\frac{1}{3}}$
and $E_F^0=(p_F^0)^2/2M$, respectively. The 
potential contribution to the energy density of symmetric nuclear matter
is denoted by $V(u)$, where $u=n/n_0$ and it is given by 
\beq
V(u)= \thalf Au^2n_0 +\frac{Bu^{\sigma+1}n_0}{1+B'u^{\sigma-1}}
+3un_0\sum_{i=1,2}C_i\left(\frac{\Lambda_i}{p_F^0}\right)^{\!3}
\left(\frac{p_F}{\Lambda_i} - \arctan\frac{p_F}{\Lambda_i}\right)\;.
\eeq
Here $p_F$ is the Fermi momentum which is related to $p_F^0$ by 
$p_F=p_F^0u^{\frac{1}{3}}$. The remaining quantities are parameters which 
are given in Refs. \cite{tpl,pal}. We choose the set which yields a 
compression modulus $K=240$ MeV for equilibrium nuclear matter since this 
is in the middle of the generally accepted 
range of $200-300$ MeV. The remaining term in Eq. (\ref{hnuc}) is the 
symmetry energy, {\it i.e.} it describes the change in the energy density 
when the proton fraction $x=n_p/n$ differs from $\thalf$.
Again following Ref. \cite{pal} we take
\beq
S(u)=\left(2^{\frac{2}{3}}-1\right)\thfive 
E_F^0\left(u^{\frac{2}{3}}-F(u)\right)+S_0F(u)\;,
\eeq
where $S_0\simeq30$ MeV is the bulk symmetry energy parameter. 
Since neutron star properties are not strongly dependent on the form
of $F(u)$, we choose the simplest of the 
suggested parametrizations namely $F(u)=u$.

In terms of the Hamiltonian density, the grand potential for nucleons is
\beq
\frac{\Omega_{N}}{\vol}={\cal H}_{N}-\mu_n n +x\mu n\;,
\eeq
where $\mu=\mu_n-\mu_p$ and the chemical potential $\mu_n$ is the Fermi energy 
for neutrons.

%
%

\subsection{Equilibrium Conditions}

We shall study cold, catalyzed neutron stars which are in chemical
equilibrium under beta decay 
processes. The process $p+e^- \leftrightarrow n+\nu_e$ in equilibrium
establishes the relation
\beq
\mu \equiv \mu_n-\mu_p = \mu_e \,, \label{beta}
\eeq
since the neutrinos have left the star and their chemical potential is 
therefore zero.
At densities where $\mu$ exceeds the muon mass $m_{\mu}$,   
muons can be formed by 
$e^-\leftrightarrow \mu^- + \overline\nu_\mu + \nu_e$, hence the muon 
chemical potential is $\mu_\mu=\mu_e=\mu$. Negatively charged kaons can be 
formed in the process $n+e^- \leftrightarrow n + K^-+\nu_e$ when 
$\mu_{K^-}$ becomes equal to the energy of the lowest eigenstate of a 
$K^-$ in matter. Chemical equilibrium requires that $\mu_{K^-}=\mu$, as we
have implicitly assumed in the text.

To complete our grand potential we need the lepton component
for which it is sufficient to use the standard non-interacting form:
\beq
\frac{\Omega_L}{V}=-\frac{\mu^4}{12\pi^2}+\eta(|\mu|-m_{\mu})\Biggl\{
\frac{m_{\mu}^4}{8\pi^2}\left[(2t^2+1)t\sqrt{t^2+1}-\ln(t+\sqrt{t^2+1})
\right] -\frac{m_{\mu}^3|\mu| t^3}{3\pi^2}\Biggr\}\,,  
\eeq
where $\eta(x)$ is the Heaviside function ($\eta(x)=1$ if $x>0$, 
$\eta(x)=0$ if $x<0$) and the quantity $t=\sqrt{\mu^2-m_{\mu}^2}/m_{\mu}$. 
The total grand potential is then
\beq
\Omega_{total} = \Omega_K + \Omega_N + \Omega_L \;.
\eeq
The values of the chemical potential $\mu$,
the proton fraction $x$ and the condensate amplitude $\theta$ are 
determined by extremizing $\Omega_{total}$.

Extremizing with respect to $\mu$ is equivalent to the requirement of 
charge neutrality. It yields
\beq
0=-n_K+n_p-n_L  \label{mumin}\;,
\eeq
where $n_K$ and $n_L$ are the number densities for negatively charged kaons 
and leptons respectively. The kaon charge density is
\beq
n_K=f^2(4b\sin^2\!\thalf\theta+\mu\sin^2\!\theta) 
-\frac{\del(V_{zp}+V_{ct})}{\del\mu} \;.
\label{nk}
\eeq
Notice that the fluctuations carry no net charge in the absence of a 
condensate, {\it i.e.} $n_K=0$ for $\theta=0$.
Here, and in the other cases, the zero point contribution requires a 
numerical integration
and it is straightforward, though tedious, to evaluate the integrand.
The lepton charge density is
\beq
n_L=\frac{\mu^3}{3\pi^2}+
\,\eta(|\mu|-m_{\mu})\frac{\mu}{|\mu|}\frac{m_{\mu}^3t^3}{3\pi^2}\;. 
\eeq
\ni
Next extremizing with respect to $x$ gives
\beq
0=n(2a_1m_s-\mu)\sin^2\!\thalf\theta +\frac{\del(V_{zp}+V_{ct})}{\del x}
+n\mu-4n(1-2x)S(u) \label{xmin}\;.
\eeq
\ni
Finally extremizing with respect to $\theta$ yields
\beq
0=f^2\sin\theta(m_K^2+c-2\mu b-\mu^2\cos\theta)
+\frac{\del(V_{zp}+V_{ct})}{\del\theta} \label{thmin}\;.
\eeq
Below the threshold for kaon condensation $\theta=0$ and this solution 
is always possible since the expression is an odd function of $\theta$.
Above threshold the solution $\theta>0$ minimizes the grand potential.
At threshold, where $\theta$ is vanishingly small, it is possible to obtain
an analytic expression by retaining terms through order $\theta^2$ in
$\Omega_K$. The extremization condition becomes
\begin{mathletters}
\label{uc}
\be
0&=& (m_K^2+c)(1+\delta)-2b\mu(1-2\delta)-\mu^2(1+\delta)
\quad {\rm where} 
\label{uca}
\ee
\be
\delta &=& \frac{1}{48\pi^2 f^2} \left( 
\xi^2-m_K^2+\xi^2\ln\frac{m_K^2}{\xi^2}\right)\;.
\label{ucb}
\ee
\end{mathletters}

This may alternatively be obtained by computing the inverse propagator,
$D^{-1}(\mu)$, to one loop order, see Appendix C. Notice that, by
construction, the inverse propagator is $(\mu^2-m_K^2)$ in vacuum 
since $b=c=\delta=0$ there. In fact for any density $\delta\leq0$. That this is
necessary may be seen as follows. The arguments of the inner square root in 
Eq. (\ref{zpdiv})
is positive for all $\bf p$, only if $c_2=c_3$ {\it or} both $c_2>0$ and 
$c_3>0$ if $c_2 \neq c_3$. However in our calculations
we have found that going even slightly above
threshold implies $c_2\neq c_3$, so that it is the latter condition that 
must be satisfied. Now the above equation (\ref{uca}) can be put 
in the form $(1+\delta)(c_2+c_3)=-6b\mu\delta$. Since
$|\delta| \ll 1$ ( Sec.III ),  
and $b$ and $\mu$ are positive, we must have $\delta<0$.
It is interesting to note that in the absence of finite counterterms 
the requirement $\delta<0$ can only be achieved in a very limited region of 
parameter space.             

Having satisfied the conditions (\ref{mumin}), (\ref{xmin}) and 
(\ref{thmin}), the energy density can be obtained from
\beq
{\cal E}=\frac{\Omega_{total}}{\vol}+\mu_n n \;.
\eeq
The chemical potential $\mu_n$ may be obtained by functional differentiation 
of the energy density with respect to the neutron density, {\it viz.}
$\mu_n=\del{\cal E}/\del n_n$, whence the pressure can be calculated 
from $P=-\Omega_{total}/\vol$. Equivalently one can write
$P=n(\del{\cal E}/\del n)-{\cal E}$.

%
%
\section{EVALUATION OF KAON LOOP CONTRIBUTION}

In this section, we give numerical estimates for the effect of the 
loop corrections upon the critical
density for kaon condensation and upon the properties of 
neutron star matter both above and below the condensate threshold.

As follows from  Eqs. (\ref{uc}) and (C1), 
the critical density for kaon condensation
is the density at which the lepton chemical potential $\mu$ (an increasing 
function of density), and the position of the 
zero-momentum $K^-$ pole of the kaon propagator
in normal nuclear matter (a decreasing function of density) become equal. 
The chemical potential $\mu$ and the proton fraction $x$, prior to 
condensation, 
are obtained from Eqs. (\ref{mumin}) and (\ref{xmin}) by setting $\theta=0$.
The pole position of the kaon propagator is a  
solution to the quadratic equation
$D^{-1}(\omega)=0$, where $D^{-1}(\omega)$ is given in Eq. (C4).

In Table I, we list the critical density for kaon condensation in units
of nuclear matter density, $u_{cr}=n_{cr}/n_0$,  
for three representative values of the parameter $a_3m_s$.
In column 2, $u_{cr}^t$ is the critical density obtained at tree 
level\cite{tpl}, 
corresponding to setting $\delta=0$ in Eq. (\ref{uc}) and omitting
the $V_{zp}+V_{ct}$ contributions from
Eqs. (\ref{mumin}) and (\ref{xmin}).
In column 3, the loop contribution is included and the corresponding
critical density is denoted by $u_{cr}^{t+l}$.
We see that $u_{cr}^{t+l}<u_{cr}^{t}$
and, while the magnitude of the threshold shift increases for
larger values of $|a_3m_s|$, it is always small.
This due to the fact that 
in Eq. (\ref{uc}) $|\delta| < m_K^2/(48\pi^2f^2) = 0.06$ which 
is small in comparison to unity. 
It is rather easy to show that $u_{cr}^{t+l}<u_{cr}^{t}$.
The tree level inverse propagator
$(D^t(\mu))^{-1}=-(c_2+c_3)=0$ at threshold, while with loops included one 
has $-(1+\delta)(c_2+c_3)=6b\mu\delta$. 
Since $D^t(\mu)$ is a monotonic function of density
and we have pointed out that $\delta<0$,
it follows that $u_{cr}^{t+l}<u_{cr}^{t}$.

The change in the threshold due to loops, small though it is, is 
dominated by the $\Sigma^{KN}$ term. This is shown in column 4 of Table I
for which we set $c=0$ in the loop expressions. In this case the threshold is 
essentially unchanged from tree level. The effect of the $c$ term can become
quite critical at densities above threshold and in Eq. (\ref{defs})
the scalar density $n_s=\langle{\bar N}N\rangle$ has been replaced by the
baryon density $n=\langle N^\dagger N\rangle$.
In Ref.\cite{lbmr} it was argued that in the heavy baryon
limit, and to the chiral order considered, 
no distinction should be made between the operators ${\bar N}N$
and $N^\dagger N$ in the Lagrangian (\ref{pje0}). Corrections to this 
would, in principle, require other effects of similar order 
for consistency. 
While $n_s\simeq 
n$ for $n\sim n_0$, at high density $n_s$ becomes less than 
$n$. The results of Ref. \cite{nsvsnb} indicate that 
this can be significant, in particular it leads to an increase in 
$u_{cr}$.  Since our results are sensitive to this question, in some of our
calculations we shall replace the baryon density 
$n$ in the quantity $c$ of Eq. (\ref{defs}) by a phenomenological form of 
$n_s$. Based on a simple parameterization of the effective nucleon mass 
obtained by Serot and Walecka\cite{sw}, we use 
\be
n_s = n \frac{2.5213}{u} \left( 1 - \frac{1}{1+\exp(\frac{u^{1/3} - 
                                        1.1717}{0.4417} ) }\right ) 
\label{ns}
\ee
for $u \ge 1.0342$ and set $n_s=n$ for  $u<1.0342$. This is extremely 
crude, but it will give some qualitative feel for the effect. The results 
for the critical density with this approach, labelled $c_{n_s}$, 
are shown in 
columns 5 and 6 of Table I. The shift in the critical density between tree 
and tree + loop levels is slightly smaller, but basically similar to that 
obtained before. The overall increase in the critical density is expected, 
but is somewhat larger than models which treat nucleons relativistically 
\cite{hyp,newborn} would suggest.

Even below threshold there is a kaon zero-point contribution 
for normal neutron star matter. Firstly there is a shift in the minimum due 
to the zero point contribution to the equilibrium Eq. (\ref{xmin}) at
$\theta=0$. Secondly there is the zero point energy (\ref{vzp0}) itself.
The shift in the proton fraction $x$ and the lepton 
chemical potential $\mu$ is negligible. Table II
shows that $\epsilon_{zp}$, the zero-point energy density is approximately
1--3\% of the total. Thus the matter properties are only affected to 
a small degree so that it is not necessary to refit the parameters of the
nuclear Hamiltonian.

Above the critical density the properties of matter  are obtained by 
solving Eqs. (\ref{mumin}), (\ref{xmin}), and (\ref{thmin}) numerically.
In practice, one uses a three-momentum cutoff $\Lambda$ to 
evaluate Eq. (\ref{zpdiv}) and derivatives thereof.
The cutoff used was $\Lambda= 3 \times 10^4$ MeV/c,
and results were found to be extremely stable with respect
to variation in $\Lambda$.  For example taking 
$\Lambda= 1 \times 10^4$ MeV/c, the largest change
in any of the tabulated values was 0.3\%.
 
In Table III, we list the properties of normal neutron star matter (column 2) 
and matter containing a kaon condensate 
at tree level (columns 3--7) and at one-loop level (columns 8--13) 
for $a_3m_s=-134$ MeV. Here
the kaon fraction is defined as $x_K=n_K/n$ in terms of the kaon density 
$n_K$ of Eq. (\ref{nk}),
and the lepton fraction $x_L=n_L/n$ may be obtained from $x_L=x-x_K$.
The energy density is also listed. We have not felt it worthwhile to 
tabulate the remaining thermodynamic variable, the pressure, in view of the 
small size of the loop effects.        
For $u\gord9.03$ the parameter $c_3$ becomes negative which precludes real 
solutions to the zero point energies; we discuss this further below.
In Table IV, we give the corresponding quantities for $a_3m_s=-222$ MeV,
in which case $c_3$ becomes negative for $u\gord5.07$.
This difficulty does not arise if we employ the phenomenological scaling given 
by Eq. (\ref{ns}) and the results are given 
for $a_3m_s=-222$ MeV in Table V. If one simply sets $c=0$ in the loop
calculations the results shown in Fig. 2 are obtained.
In this figure $\Delta \epsilon$ represents
the energy density gained by the formation of a condensate and is 
the difference between the total energy density and the tree level energy
density for normal nuclear matter.  

We first note that loop effects for the case $a_3m_s=-134$ MeV
are essentially negligible. Although they increase in magnitude for 
larger values of $|a_3m_s|$, only small modifications
of the tree level results are found. Table IV shows that just above 
threshold the tree + loop calculation gives values of
$\theta$ and $x$ which are greater than, and
$\mu$ which are less than, the corresponding tree level results.
This naturally follows from the fact that $u_{cr}^{t+l}$ is less than 
$u_{cr}^{t}$ since, once a condensate is present, $\theta$ rises rapidly and
$x$ is enhanced, while $\mu$ is reduced 
At high densities, this ordering is reversed.
Near the limiting density, $u\simeq5.07$,
$\theta$  typically decreases by $\sim3$\%
while $x$ decreases somewhat less, $\sim1$\%, and $\mu$ increases
$\sim6$\%.
The zero-point energy density is of the order 
$-(1.2-3.5) \, {\rm MeV} \,{\rm fm}^{-3}$
and can decrease the tree level result significantly at high density when 
the latter can be small. For a given density the effects are 
smaller if the scalar density of Eq. (\ref{ns}) is employed as shown in 
Table V. On the other hand, setting $c=0$ altogether as in Fig. 2 yields 
larger effects at very high density.
For $a_3m_s=-310$ MeV, the above trends are also present, and are 
enhanced, as expected.
The coefficient $c_3$ goes to zero for $u\gord3.32$,
at which point $\epsilon_{zp} \simeq -7.2 \,{\rm MeV}\,{\rm fm}^{-3}$
and $\epsilon \simeq -8.5 \,{\rm MeV}\,{\rm fm}^{-3}$. 

As mentioned above,
$c_3$ often vanishes at some density above the critical 
density.  The condition $c_3=0$ implies $d\epsilon/d\theta=0$,
where $\epsilon$ is the tree level energy density function
of Eq. (\ref{vtree}).
The significance of this condition is best illustrated by the simple
$O(2)$ invariant Lagrangian ${\cal L}= \frac12 
( \del_\nu \phi_1 )^2 + \frac12 ( \del_\nu \phi_2 )^2 - U(\phi)$,
where $\phi^2=\phi_1^2 + \phi_2^2$.
The one-loop effective potential\cite{pokorski} is 
$U^{l}(\phi) = i \tr \ln (( k^2 + U^{\prime \prime})
(k^2 + U^\prime/\phi))$, where the trace is over momentum.
Now take $U(\phi)=\frac12 m^2 \phi^2 + \lambda \phi^4$,
with $\lambda>0$. For $m^2>0$, $U^{l}$ is real for all 
$\phi$.  On the other hand, for $m^2<0$, the ``Mexican hat''
potential, the one-loop effective potential is complex
in the region $U^\prime<0$ which signals a physical instability.
Weinberg and Wu \cite{wwu} show that the imaginary part can be 
interpreted as a decay rate per unit volume of a well-defined state.
The connection between the instability and the chaotic dynamics of a 
classical field theory has recently been discussed by 
Matinyan and M\"uller \cite{mm}, see also references therein.         

Unfortunately no solution is known to this general problem.
Our case is entirely analogous. At threshold $c_2=c_3>0$ corresponds
to $m^2>0$.  As the density is increased, the tree level potential
eventually corresponds to $m^2<0$.  This effect is more pronounced
for larger values of $|a_3m_s|$.  Since the effective potential is 
by definition real, this shows that some, as yet unknown, technique 
is needed in these regions of parameter space. 
The two other cases we have considered, excluding $c$ 
from the loop contribution and employing the softer density 
dependence of $c$ with the parametrized scalar density $n_s$
alleviate this problem.

%
%

\section{Conclusion}

We have derived the zero-point energy due to fluctuations in kaonic
modes in neutron star matter containing a kaon condensate. 
To our knowledge, this calculation is the first in which the nonlinearities 
of the chiral Lagrangian have been
taken into account at the one loop level for densities
{\it above} the critical density for kaon condensation.
The zero-point energy was derived through the thermodynamic
partition function by considering fluctuations around an
arbitrary condensate.
We have demonstrated by gauging the counterterms
of the vacuum theory that no divergences are present when matter is 
introduced via the charge chemical potential.
In the usual way the total thermodynamic
potential, including the classical contribution, is extremized so that 
the minimum energy ground state is found for locally charge neutral and 
beta-equilibrated matter.

The critical density for kaon condensation was found to be reduced
by fluctuations, but only by less than 1\% for the range of parameters 
considered. At higher densities the loop effects were negligible 
for $a_3m_s=-134$ MeV which corresponds to zero strangeness content for 
the nucleon. The effects were slightly increased with larger
values of $|a_3m_s|$, corresponding to larger value of the kaon
nucleon sigma term $\Sigma^{KN}$, but they remained small in comparison to 
the dominant tree level contributions. 
The size of the loop effects relative to the tree contributions is set, at
least in the threshold region, by the parameter $\delta$ which is less than
$m_K^2/(48\pi^2 f^2)=0.06$. 
In some cases it was not possible to
obtain solutions up to arbitrary density because the effective potential 
became complex-- this is a well-known and unsolved problem in one loop 
calculations with spontaneous symmetry breaking.

The fact that changes induced by the zero-point
motion are generally small leads one first to conclude that no major revision 
is needed of earlier investigations in which such fluctuations
have not been taken into account.  Second, the loop expansion
appears to be convergent where it is well defined. 
The formalism presented here yielded the contribution to the grand 
potential arising from thermal excitations of the fluctuation modes.
These effects would have to be included at finite temperatures
and would be involved in the extremization conditions.
Such an analysis is needed for a consistent study 
of newborn neutron stars.         
Extension of the present work to include the full
octet of pseudoscalar fluctuations, and to include
additional terms in the chiral expansion 
could also be contemplated.

%
%

\acknowledgments

We wish to thank P. B. Arnold, 
A. Fayyazuddin, U. van Kolck, and L. Yaffe for helpful discussions,
and D. B. Kaplan, M. Rho, R. Venugopalan and A. Wirzba for
valuable comments on the manuscript.  
We are grateful to M. Prakash for stimulating our interest in 
this problem and for careful reading of the manuscript. 
This work was supported in part by the U. S. Department of 
Energy under Grant DE-FG06-88ER40427.

\appendix

%
%

\section{ Cancellation of chemical potential divergences in a toy model}

In this Appendix,   
we illustrate how divergences dependent on the 
charge chemical potential $\mu$ are cancelled in 
a toy model which contains two real fields, $\phi_1$and $\phi_2$.
The model we consider is  
\be
{\cal L} = {\cal L}_\gamma + {\cal L}_{ct} \quad{\rm where}\quad
{\cal L}_\gamma
= \thalf\gamma \left[ ( \del_\nu \phi_1 )^2 + ( \del_\nu \phi_2 )^2 
\right]\;, 
\label{gtoytree}
\ee
and $\gamma$ is an arbitrary analytic 
function of $\phi^2 = ( \phi_1^2+\phi_2^2 )$
with $\gamma(0)=1$.  It is useful to discuss this model to one loop order
since it has a simpler structure than the chiral Lagrangian,
but the $\mu$-dependent divergences are removed in a similar way.
We first derive ${\cal L}_{ct}$,
and then demonstrate that on introducing 
an external chemical potential $\mu$, the  
effective potential (or equivalently,
the zero-point energy) is free of divergences. Finite counterterms are not
relevant in the present context and will be ignored.        

Expanding the function $\gamma$ to all orders in the fields
gives the the standard kinetic energy term plus
an infinite set of interaction terms involving
an even number of fields and two derivatives. 
{}From these interaction vertices, we may compute 
one-particle irreducible $2n$-point
Green functions to one-loop order.  
The counterterm Lagrangian ${\cal L}_{ct}$ is
defined in such a way as to minimally cancel divergences. 

Consider first quadratic divergences, proportional to $d_2$ 
(defined in Eq. (\ref{quaddiv})). 
The 4-point vertex contributes a quadratic
divergence to the 2-point function proportional to $d_2$
and the external momenta.  
The corresponding counterterm Lagrangian is proportional
to $d_2[(\del_\nu \phi_1)^2+ (\del_\nu \phi_2)^2]$. 
The 6-point vertex contributes to the 4-point function
so as to require a counterterm Lagrangian proportional 
to $d_2 \phi^2 [(\del_\nu \phi_1)^2+ (\del_\nu \phi_2)^2]$.
Continuing this sequence, we find that the contribution of 
quadratic divergences to the counterterm Lagrangian is 
$-d_2/(4\gamma) \cdot (\gamma''+\gamma'/\phi)
[(\del_\nu \phi_1)^2+ (\del_\nu \phi_2)^2]$,
where primes denote derivatives with respect to $\phi$.

Now consider logarithmic divergences, proportional to $d_0$
(see Eq. (\ref{logdiv})).  Green functions containing such
divergences typically contain two vertices, and 
evaluating them requires involved combinatorics. It is simpler instead 
to derive the counterterm Lagrangian using the effective potential
(see e.g. Ref.\cite{pokorski}).
One thus avoids combinatorics, and the method is 
more easily applied to Lagrangians more complicated
than Eq. (\ref{gtoytree}).

Let
\be
\phi_1 \rightarrow \phi_1 + \phi_1'   \,\,\, , \,\,\phi_2 
\rightarrow \phi_2 + \phi_2'  \,\,\, ,
\label{phiexpand}
\ee
where $\phi_1$ and $\phi_2$ are arbitrary fields which, for the present,
may vary in spacetime.  Expanding to quadratic 
order in the fields $\phi_1'$ and $\phi_2'$ leads to 
\be
{\cal L}_\gamma  \rightarrow {\cal L}_\gamma +
\thalf
(
\begin{array}{cc}
\phi_1' & \phi_2'
\end{array}
)
D
\left(
\begin{array}{c}
\phi_1' \\ \phi_2'
\end{array}
\right)
\,\,\,,
\ee
where the matrix of second order variations is given by 
\be
D &=& - \gamma \Box\vm{1} + L \cdot \del - M \,\,\, , \\
L_\nu &=& 
2 \frac{\gamma'}{\phi} 
\left(
\begin{array}{cc}
\phi_1 \del_\nu \phi_1  & \phi_1 \del_\nu \phi_2\\
\phi_2 \del_\nu \phi_1 &  \phi_2 \del_\nu \phi_2
\end{array}
\right) \,\,\, , \\
M &=& - 
\frac{1}{2 \phi^2} \left[  (\del_\nu \phi_1)^2 +  (\del_\nu \phi_2)^2 
\right]
\left(
\begin{array}{cc}
\phi_1^2 \gamma'' + \phi_2^2 \frac{\gamma'}{\phi}  
& \phi_1 \phi_2 ( \gamma'' - \frac{\gamma'}{\phi} ) \\
 \phi_1 \phi_2 ( \gamma'' - \frac{\gamma'}{\phi} ) 
& \phi_2^2 \gamma'' + \phi_1^2 \frac{\gamma'}{\phi}
\end{array}
\right)
\,\,\,.
\ee
Integrating out the fields $\phi_1'$ and $\phi_2'$ gives the
unrenormalized effective potential to one-loop order 
\be
-\frac{i}{2}  \tr \ln D      
&=& - i \tr \ln \gamma -i \tr \ln (-\Box)  \nonumber\\ 
&& + \frac{i}{2}  \sum_{k=1}^{\infty} \frac{1}{k} 
\tr \left( \frac{1}{-2\gamma \Box} 
[\stackrel{\leftarrow}{\del} \cdot L -
L \cdot \stackrel{\rightarrow}{\del} 
+ 2M + (\del \cdot L) ] \right)^k  
\,\,\,,
\label{effpotgamma}
\ee
where, as before, the trace is taken both over spacetime  and charge. 
The factor $- i \tr \ln \gamma$ is cancelled by 
a corresponding contribution from the measure, 
and will not  be considered further. 
In the last expression, we have symmetrized the derivative contribution 
and separated the mass-like contributions involving 
$[2M + (\del \cdot L)]$.  We shall see that only the mass-like  
contributions are important for our discussion. 

The trace over spacetime is evaluated by inserting complete
sets of four-momentum eigenstates.  A momentum expansion
is performed on the resulting integrals, leading
to a factorization into ultraviolet divergent loop integrals
and integrals over functions of the fields. 
To minimally cancel the divergences in the effective potential
we need the following counterterm Lagrangian
\be
{\cal L}_{ct} =  {\cal L}_{ct}^{quartic}
+ {\cal L}_{ct}^{mass-like} + {\cal L}_{ct}^{derivative}  \,\,\,,
\ee
where the contribution from the quartic divergence (\ref{quartdiv}) is
\be
\int d^4x\:{\cal L}_{ct}^{quartic} = -i \tr \ln ( - \Box ) = \int d^4x\:  d_4
\label{toyquart}\,\,\,.
\ee
The contribution from mass-like terms is 
\be
\int d^4x \:{\cal L}_{ct}^{mass-like} 
&=& \frac{i}{2}  \sum_{k=1}^{2} \frac{1}{k} 
\tr \left( \frac{1}{-2\gamma \Box} 
[2M + (\del \cdot L) ]\right)^k \nonumber \\
&=& \frac{d_2}{4 \gamma} \int d^4x \:\tr [2M + (\del \cdot L) ]
- \frac{d_0}{16\gamma^2} \int d^4x \:\tr 
[ 2M  + (\del \cdot L)]^2  \nonumber \\
&=& - \frac{d_2}{4 \gamma} \int d^4x 
\:\Biggl\{\left( \gamma'' + \frac{\gamma'}{\phi} \right)
\left[  (\del_\nu \phi_1)^2 +  (\del_\nu \phi_2)^2 \right] 
-\frac{\gamma'}{\phi}\Box\phi^2\Biggr\}\nonumber\\
&&-\frac{d_0}{4\gamma^2} \int d^4x 
\:\Biggl\{\left[ (\del_\nu \phi_1)^2 +  (\del_\nu \phi_2)^2 \right]^2
\Biggl[ \left(\frac{\gamma''}{2}\right)^2+
\left(\frac{\gamma'}{2\phi}\right)^2 \Biggr]
 \nonumber\\
&& - \frac{\gamma'}{\phi^3} \left[ (\del_\nu \phi_1)^2 +  (\del_\nu \phi_2)^2 
\right]\biggl[ \gamma'' ( \phi_1 \del_\nu \phi_1 
+ \phi_2 \del_\nu \phi_2 )^2\nonumber\\ 
&& +\gamma''\phi^2 ( \phi_1 \Box \phi_1 + \phi_2 \Box \phi_2 )
+\frac{\gamma'}{\phi} ( \phi_1 \del_\nu \phi_2 -  
\phi_2 \del_\nu \phi_1 )^2 \biggr]\nonumber\\ 
&& + \left(\frac{\gamma'}{\phi}\right)^{\!2} \biggl[ 
( \del_\nu\phi_1 \del_\lambda \phi_1 
+ \del_\nu\phi_2 \del_\lambda\phi_2 )^2 
+ (\phi_1 \Box \phi_1 + \phi_2 \Box \phi_2 )^2 \nonumber\\
&& +  2\Bigl( (\del_\nu \phi_1)^2 \phi_1 \Box \phi_1 
+ (\del^\nu \phi_1 \del_\nu \phi_2) \phi_1 \Box \phi_2 
+ \{1 \leftrightarrow 2 \}  \Bigr)   
\biggr] \Biggr\}\;.
\label{masslike}
\ee
For the contribution from derivatives one finds (see also Refs. \cite{gl,zwm})
\be
\int d^4x \:{\cal L}_{ct}^{derivative} &=& 
\frac{i}{2} \sum_{k=2}^{4} \frac{1}{k} 
\tr \left( \frac{1}{2 \gamma \Box} 
( L \cdot \stackrel{\rightarrow}{\del} 
- \stackrel{\leftarrow}{\del} \cdot L ) \right)^k    \nonumber\\
&=& - \frac{d_0}{96} \int d^4x 
\:\tr\Biggl\{ \frac{1}{\gamma^2} ( \del_\nu L_\lambda -
\del_\lambda L_\nu )^2 
- \frac{1}{\gamma^3} ( \del_\nu L_\lambda -
\del_\lambda L_\nu ) [ L_\lambda, L_\nu] \nonumber \\
&& + \frac{1}{\gamma^4} L_\nu^2
L_\lambda^2 + \frac{1}{2\gamma^4}  (L_\nu L_\lambda)^2 \Biggr\} 
- \frac{d_2}{8 \gamma^2} 
\int d^4x\: \tr L_\nu^2
\,\,\,.
\label{dervs}
\ee
This concludes the derivation of ${\cal L}_{ct}$, to one-loop
order.

Having derived a one-loop finite vacuum theory, 
we are now in a position to study  
the theory in the presence of a finite
chemical potential $\mu$, at finite inverse temperature $\beta=1/T$.
As before, the Euclidean time is $\tau = i t$, 
and the action $e^{iS} \rightarrow e^{-S}$ under the transformation
to Euclidean space. 
The chemical potential acts as an external vector field with
spatial components zero. It enters by ``gauging'' the 
fields $\phi_1$ and $\phi_2$ via the transformation
\be
i \frac{\del \phi_1}{\del \tau} \rightarrow 
i \frac{\del \phi_1}{\del \tau} - \mu \phi_2  \quad;\quad
i \frac{\del \phi_2}{\del \tau} \rightarrow 
i \frac{\del \phi_2}{\del \tau} + \mu \phi_1   \,\,\,.
\label{imtrans}  
\ee
The resulting action is given by 
\be S &=& S_\gamma + S_{ct}    \quad,\ {\rm where}
\label{sgammamu}
\\
S_\gamma
&=& \thalf \int\limits_0^\beta d\tau \int d^3x \:\gamma {\cal L}_\mu \quad,
\quad
S_{ct} =
\int\limits_0^\beta d\tau \int d^3x \:{\cal L}_{ct} \quad{\rm and}  \\ 
{\cal L}_{ct}  
&=& - d_4 
- \frac{d_2}{4 \gamma} 
\left\{\left( \gamma'' + \frac{\gamma}{\phi} \right) {\cal 
L}_\mu+\cdots\right\}
+ \frac{d_0}{16 \gamma^2} \left\{
\left[(\gamma'')^2+\left(\frac{\gamma'}{\phi}\right)^2\right]
({\cal L}_\mu)^2+\cdots\right\} \,\,\, ,\label{toylct}\\   
{\cal L_\mu} &=& ( \del_\tau \phi_1 )^2 + ( \del_\tau \phi_2 )^2 
+ ( \bbox{\nabla}\phi_1 )^2 +   
( \bbox{\nabla}\phi_2 )^2 +2 i \mu ( \phi_2 \del_\tau
\phi_1 - \phi_1 \del_\tau \phi_2 ) - \mu^2 \phi^2
\,\,\,.
\ee
Eq. (\ref{dervs}) is invariant under the
transformation (\ref{imtrans}) and is therefore not included above. 
We have now derived the action to one-loop order,
for arbitrary fields $\phi_1$ and $\phi_2$ in the presence
of a chemical potential $\mu$. 

We then take the final step, 
and evaluate the one-loop effective potential for fields $\phi_1$ and $\phi_2$ 
which are {\it constant} in spacetime (see Eq. (\ref{decomp})) so that their 
derivatives vanish.  
Expanding Eq. (\ref{sgammamu}) to quadratic order 
in the fields and integrating out the fluctuating field
gives the action  
\be
S &=& S_{tree} + S_{loop} + S_{ct}  \,\,,
\nonumber \\
S_{tree} &=& - \thalf\int\limits_0^\beta d\tau \int d^3x \:\gamma \mu^2 \phi^2
\,\,\,,
\\
S_{loop} &=& -\thalf\ln {\rm det} D  \,\,\,,
\label{sloop}
\ee
where the matrix of second order variations about the constant
fields, $D$, is given by 
\be
\frac{1}{\beta^2} D_{11}  &=&
\gamma( \omega_n^2 + {\bf p}^2 ) 
- 2 \omega_n \mu \phi_1 \phi_2 \frac{\gamma'}{\phi}
- \mu^2 \gamma 
- \thalf\mu^2 \frac{\gamma'}{\phi} ( \phi_2^2 + 4 \phi_1^2 )
- \thalf\mu^2 \phi_1^2 \gamma''    \nonumber\\
\frac{1}{\beta^2} D_{22}  &=& 
\gamma( \omega_n^2 + {\bf p}^2 ) 
+ 2 \omega_n \mu \phi_1 \phi_2 \frac{\gamma'}{\phi}
- \mu^2 \gamma 
- \thalf\mu^2 \frac{\gamma'}{\phi} ( \phi_1^2 + 4 \phi_2^2 )
- \thalf\mu^2 \phi_2^2 \gamma''    
\nonumber \\
\frac{1}{\beta^2} D_{12}  &=&
2 \mu \omega_n \left( \gamma + \phi_1^2 \frac{\gamma'}{\phi} \right)
- \thalf \mu^2 \phi_1 \phi_2 \left( \gamma'' + 3 \frac{\gamma'}{\phi} 
\right) \nonumber\\
\frac{1}{\beta^2} D_{21}  &=&
- 2 \mu \omega_n \left( \gamma + \phi_2^2 \frac{\gamma'}{\phi} \right)
- \thalf\mu^2 \phi_1 \phi_2 \left( \gamma'' + 3 \frac{\gamma'}{\phi} \right)
\,\,,
\ee
and $\omega_n$ is the Matsubara frequency. 
The determinant is given by Eq. (\ref{detD}), with an additional
factor of $\gamma^{-1}$ on the left, and with 
\be
c_1 = \frac{2 \mu^2}{\gamma} ( \gamma + \phi \gamma' )
\ ,\ 
c_2 = - \frac{\mu^2}{4 \gamma} ( 2 \gamma + 4 \gamma'\phi + \phi^2 \gamma'')
\ ,\ 
c_3 = - \frac{\mu^2}{4 \gamma} ( 2 \gamma + \gamma'\phi ) \,\,\,.
\ee
As before, it is convenient to factorize the determinant, leading to 
dispersion relations for two modes given by 
Eq. (\ref{disp}).  Eq. (\ref{sloop}) then separates into a zero-point
and a thermal contribution 
\be
S_{loop}  = \int\limits_0^\beta d\tau \int d^3x \:( V_{zp} + V_{thermal})\;,
\ee
with $V_{zp}$ and $V_{thermal}$ given by Eqs. (\ref{vzp}) and 
(\ref{vthermal}) respectively. 
The counterterm action obtained from Eqs. (\ref{toyquart}), (\ref{masslike})
and (\ref{dervs}) after gauging and setting $\phi_i$ to
be constant can be written
\be
S_{ct}  = \int\limits_0^\beta d\tau \int d^3x  \:V_{ct}  \,\,\,,     
\ee
where $V_{ct}$ is given by the divergent parts of Eq. (\ref{ct}), {\it 
i.e.} the terms involving $d_0$, $d_2$ and $d_4$. 
As noted in Sec.
IIB, the $\mu$ dependent divergences in $V_{zp}$ are exactly cancelled
by those in $V_{ct}$.

%
%

\section{ Cancellation of chemical potential divergences in the chiral
		Lagrangian}

In this appendix, we give details of the cancellation of 
quadratic divergences dependent on the chemical
potential $\mu$ for the chiral Lagrangian, Eq. (\ref{lphi}).
(The cancellation of quartic divergences needs no discussion since it 
is straightforward.) We follow the procedure given in Appendix A.
It is first convenient to break the Lagrangian down as follows:
\be
{\cal L}_K &=& {\cal L}_H+ {\cal L}_V + {\cal L}_U
+ {\cal L}_{ct} \nonumber\\
{\cal L}_H
&=& \thalf \, A  \, (\del_\nu \phi_1)^2
+ \thalf \, B  \, (\del_\nu \phi_2)^2
+ C \, \del_\nu \phi_1 \del_\nu \phi_2 \nonumber \\
{\cal L}_V &=& b \gamma_\frac12
(\phi_1\stackrel{\leftrightarrow}{\del^0}\phi_2)  \nonumber \\
{\cal L}_U &=& - 2 f^2 (m_K^2+c) \sin^2\!\thalf\Theta \equiv - U(\phi)\;.
\ee
Note that in addition to having more terms than the toy model
of Appendix A, the kinetic term, ${\cal L}_H$, is nondiagonal here. 

Expanding as in Eq. (\ref{phiexpand}), the matrix $D$ of second 
order variations of the fields $\phi_i'$ is given by 
\beq
D = -T \Box + L \cdot \del - M   \quad;\quad
T = 
\left(
\begin{array}{cc}
A & C \\ 
C & B
\end{array}
\right)\;.
\eeq
The matrix $L$ can be written
\be
L^\nu &=& L_{H_1}^\nu + L_{H_2}^\nu   + L_{V_1}^\nu + L_{V_2}^\nu 
\qquad{\rm where}\nonumber\\
L_{H_1}^\nu  &=& 2 \frac{\gamma'}{\phi} 
\left(
\begin{array}{cc}
\phi_1 \del^\nu \phi_1  & \phi_1 \del^\nu \phi_2\\
\phi_2 \del^\nu \phi_1 &  \phi_2 \del^\nu \phi_2
\end{array}
\right) 
\quad;\quad
L_{H_2}^\nu = 
- 2 \frac{1-\gamma}{\phi^2} 
\left(
\begin{array}{cc}
\phi_1 \del^\nu \phi_1  & \phi_1 \del^\nu \phi_2\\
\phi_2 \del^\nu \phi_1 &  \phi_2 \del^\nu \phi_2
\end{array}
\right) 
\nonumber \\
L_{V_1}^\nu &=& \delta^{\nu 0} 2 b \gamma_\frac12
\left(
\begin{array}{cc}
0 & 1 \\ 
-1 & 0
\end{array}
\right)
\quad;\quad
L_{V_2}^\nu =
\delta^{\nu 0} 2 b \frac{\gamma_\frac12'}{\phi}
\left(
\begin{array}{cc}
- \phi_1 \phi_2 &  \phi_1^2 \\ 
- \phi_2^2  & \phi_1 \phi_2 
\end{array}
\right) \;. \label{lterms}
\ee
The matrix $M$ takes the form
\be
M &=& M_{H_1} + M_{H_2} + M_{V_1} + M_{V_2} + M_U \quad{\rm where}\nonumber \\
M_{H_1} &=& -  
\frac{1}{2 \phi^2} \left[ (\del_\nu \phi_1)^2 + (\del_\nu \phi_2)^2 \right]
\left(
\begin{array}{cc}
\phi_1^2 \gamma'' + \phi_2^2 \frac{\gamma'}{\phi}  
& \phi_1 \phi_2 ( \gamma'' - \frac{\gamma'}{\phi} ) \\
 \phi_1 \phi_2 ( \gamma'' - \frac{\gamma'}{\phi} ) 
& \phi_2^2 \gamma'' + \phi_1^2 \frac{\gamma'}{\phi}
\end{array}
\right)  \nonumber \\
M_{H_2} &=& 
\frac{1-\gamma}{\phi^2}
\left(
\begin{array}{cc}
\phi_1 \Box \phi_1  & \phi_1 \Box \phi_2\\
\phi_2 \Box \phi_1 &  \phi_2 \Box \phi_2
\end{array}
\right)
\nonumber \\
M_{V_1} &=& 2 b \frac{\gamma_\frac12'}{\phi}
\left(
\begin{array}{cc}
- \phi_1 \del^0 \phi_2  & - \phi_2 \del^0 \phi_2\\
\phi_1 \del^0 \phi_1 &  \phi_2 \del^0 \phi_1
\end{array}
\right) \nonumber \\
M_{V_2} &=&
- b (\phi_1\!\stackrel{\leftrightarrow}{\del^0}\!\phi_2)
\frac{1}{\phi^2}
\left(
\begin{array}{cc}
\phi_1^2 \gamma_\frac12'' + \frac{\phi_2^2}{\phi} \gamma_\frac12'
& \phi_1 \phi_2 ( \gamma_\frac12'' - \frac{1}{\phi}\gamma_\frac12') \\
 \phi_1 \phi_2 ( \gamma_\frac12'' - \frac{1}{\phi}\gamma_\frac12' ) 
& \phi_2^2 \gamma_\frac12'' + \frac{\phi_1^2}{\phi}\gamma_\frac12'
\end{array}
\right) \nonumber \\
M_U &=&  
\frac{1}{\phi^2}
\left(
\begin{array}{cc}
\phi_1^2 U'' + \phi_2^2 \frac{U'}{\phi}  
& \phi_1 \phi_2 ( U'' - \frac{U'}{\phi} ) \\
 \phi_1 \phi_2 ( U'' - \frac{U'}{\phi} ) 
& \phi_2^2 U'' + \phi_1^2 \frac{U'}{\phi}
\end{array}
\right) \;. \label{mterms}
\ee
As in Appendix A primes denote derivatives with respect to $\phi$.
It may be checked that by gauging $D$ according to Eq. (\ref{imtrans})
and then taking the fields $\phi_i$ to
be constant, one recovers Eq. (\ref{Dmatrix}).

It is now straightforward, but tedious, to compute the 
the quadratically divergent contributions to the 
counterterm Lagrangian by evaluating the terms of the form
\be
\thalf \tr  ( T^{-1} M)  - \teighth \tr ( T^{-1} L^\nu )^2 \,\,.
\ee
Omitting terms that are invariant under the transformation (\ref{imtrans}),
we find the following contributions to ${\cal L}_{ct}/d_2$:
\be
\thalf \tr  ( T^{-1} M_{H_1}) &=&
- \tquar
\left[ (\del^\nu \phi_1)^2 + (\del^\nu \phi_2)^2 \right]
\left( \gamma'' +  \frac{\gamma'}{\gamma\phi} \right)
\label{mh1}
\\
\thalf\tr  ( T^{-1} M_{H_2}) &=&
\frac{ 1-\gamma }{2 \phi^2 } ( \phi_1 \Box \phi_1 + \phi_2 \Box \phi_2 )
\label{mh2}
\\
\thalf \tr  ( T^{-1} M_{V_1}) &=&
- b \frac{\gamma_\frac12'}{\phi} 
\,( \phi_1\stackrel{\leftrightarrow}{\del^0}\phi_2 )
\label{mv1}
\\
\thalf \tr  ( T^{-1} M_{V_2}) &=&
- \thalf b 
\left( \gamma_\frac12'' + \frac{\gamma_\frac12'}{\gamma\phi} \right)
( \phi_1\stackrel{\leftrightarrow}{\del^0}\phi_2 )
\label{mv2}
\\
\thalf \tr  ( T^{-1} M_U) &=&
\thalf \left(  \frac{U'}{\gamma\phi} + U'' \right)  
\label{mu}
\\
- \teighth \tr ( T^{-1} L_{V}^\nu )^2 &=&
b^2 \frac{\gamma_\frac12}{\gamma}
( \gamma_\frac12 + \phi\gamma_\frac12')
\label{lv}
\\
- \teighth \tr ( T^{-1} [ L_{H_1}^\nu + L_{H_2}^\nu ] )^2 &=&
- \thalf \left( \frac{\gamma'}{\phi} - \frac{1-\gamma}{\phi^2}\right)^{\!2} 
( \phi_1 \del^\nu \phi_1 + \phi_2 \del^\nu \phi_2 )^2
\label{lhsq}
\\
- \tquar\tr ( T^{-1} L_{V}^\nu T^{-1} 
[ L_{H_1}^\nu + L_{H_2}^\nu ] ) &=&
b \frac{\gamma_\frac12}{\gamma} \left( \gd - \frac{1-\gamma}{\phi^2} \right)
( \phi_1\stackrel{\leftrightarrow}{\del^0}\phi_2 )
\label{lvlh}       \,\,\,.
\ee

Following Appendix A, we use the transformation (\ref{imtrans})
on ${\cal L}_{ct}$ and find that the quadratically divergent
contribution to $V_{ct}$ is
\be
V_{ct}^{quad} &=& d_2 
\Biggl\{ \tquar \mu^2 \phi^2 \left( \gdd + \frac{\gamma'}{\gamma\phi} 
\right) 
+ \thalf \mu^2 ( 1 - \gamma )
+ b \mu \phi \gamma_\frac12'
+ \thalf b \mu \phi^2 \left( \gtdd + \frac{\gamma_\frac12'}{\gamma\phi} 
\right) \nonumber\\ 
& & +b\mu\frac{\gamma_\frac12}{\gamma} ( 1-\gamma - \phi\gamma') 
- b^2 \frac{\gamma_\frac12}{\gamma} ( \gamma_\frac12 + \phi\gamma_\frac12')
- \thalf \left(U'' + \frac{U'}{\gamma\phi} \right) \Biggr\}\;.
\ee
It may be checked that this agrees with the divergent parts of
Eq. (\ref{ct}) using
the explicit form of the coefficients $c_1$, $c_2$ and $c_3$:
\be
c_1 
&=& 2 \mu^2 ( \gamma + \phi\gamma' ) 
+ 2 \mu b \left( 2 \gamma_\frac12 + 
\frac{\phi\gamma'\gamma_\frac12}{\gamma} + \phi\gamma_\frac12'\right)
+ 2 b^2 \frac{\gamma_\frac12}{\gamma} ( \gamma_\frac12 
+ \phi\gamma_\frac12')
\\
c_2 &=& 
- \tquar \mu^2 ( 2\gamma + 4\phi\gamma' + \phi^2 \gamma'') 
- \thalf \mu b ( 2 \gamma_\frac12 + 4 \phi\gamma_\frac12' + 
\phi^2\gamma_\frac12'') +\thalf U'' 
\\
c_3 &=&   
- \tquar \mu^2 \left( 2 + \frac{\phi\gamma'}{\gamma} \right)
- \thalf \mu b  \frac{1}{\gamma}( 2\gamma_\frac12 + \phi\gamma_\frac12')   
+\frac{U'}{2\gamma\phi}  \,\,\,.
\ee

We now discuss the chiral ordering of the above counterterms
in the Lagrangian (\ref{pje0}).
To do so, it helpful to evaluate the divergent integrals of Eqs.
(\ref{quartdiv}), (\ref{quaddiv}) and (\ref{logdiv}) using 
dimensional regularization. This gives
\be
d_4 &=& \int \frac{d^3k}{(2\pi)^3} \sqrt{k^2+m_K^2} 
= 
-\frac{m_K^4}{16\pi^2}
\left( \frac{1}{\epsilon}+\ln\frac{\kappa}{m_K}+\thalf\Psi(3)
\right)
\nonumber \\
d_2 &=& \int \frac{d^3k}{(2\pi)^3} \frac{1}{2\sqrt{k^2+m_K^2}}
= 
-\frac{m_K^2}{8\pi^2}
\left( \frac{1}{\epsilon}+\ln\frac{\kappa}{m_K}+\thalf\Psi(2)
\right)
\nonumber \\
d_0 &=& \int \frac{d^3k}{(2\pi)^3} \frac{1}{4(k^2+m_K^2)^{3/2}} 
=
\frac{1}{8\pi^2}
\left( \frac{1}{\epsilon}+\ln\frac{\kappa}{m_K}+\thalf\Psi(1)
\right)
\;. 
\label{dimregds}
\ee
Here, we have included $m_K$ explicitly in the kaon propagator.
A factor of $\sqrt{4\pi}$ has been subsumed in $\kappa$ which 
is introduced so that the dimension of the integral
is independent of the dimensionality of spacetime.
The parameter $\epsilon=3-n$, where
$n$ is the dimensionality of the integral, and $\Psi(n)$ is the digamma 
function \cite{as}.
In minimal subtraction the integrals of Eq. (\ref{dimregds})
are regulated by removing the $1/\epsilon$ terms. 
It is then evident that in the chiral expansion
$d_4$, $d_2$ and $d_0$ are ${\cal O}(q^4)$, ${\cal O}(q^2)$
and ${\cal O}(q^0)$ respectively. 

To evaluate the chiral order of counterterms, 
we promote the nucleon fields in $b$ to dynamical fields.
We may then determine the chiral order of 
counterterms from Weinberg's counting rule (\ref{count}), 
either directly from the tree-level counterterms
or from the representation of these
terms as one-loop diagrams. 
Counterterms (\ref{mv1}), (\ref{mv2}) and (\ref{lvlh})
contribute to ${\cal L}^{(3)}_{ct}$.
Expanding to lowest order in the fields $\phi_1$ and $\phi_2$,
these terms  renormalize the kaon-nucleon vector interaction
coupling constant.
Counterterms (\ref{mh1}), (\ref{mh2}) and (\ref{lhsq})
contribute to ${\cal L}^{(4)}_{ct}$.
To lowest order in the fields, they
contribute to wave function renormalization.  
Eq. (\ref{mu}) is also ${\cal O}(q^4)$, but to 
lowest order in the fields gives an overall irrelevant
constant and a mass renormalization. 
Eq. (\ref{lv}) contributes to ${\cal L}^{(2)}_{ct}$ and yields
a four-nucleon interaction to lowest order in the 
fields. 

In this Appendix, we have not discussed logarithmic divergences.
We believe the cancellation of $\mu$ dependent logarithmic divergences
will proceed as in the toy model of Appendix A.
However we have not derived the full counterterm Lagrangian proportional
to $d_0$ since this, while in principle straightforward,
would involve prohibitively complicated algebra.
The counterterms will have a chiral order which ranges from 0 to 4 as
can be seen from Eqs. (\ref{masslike}) and (\ref{dervs}) in conjunction
with Eqs. (\ref{lterms}) and (\ref{mterms}).
In conclusion we note that although we have referred to
chiral counting throughout this manuscript,   
our Lagrangian should be regarded as a ``U(1)'' or ``O(2)''
analogue of the chiral expansion.

%
%

\section{Critical density from the kaon propagator and from expansion
of zero-point energy}

In this appendix, we derive the critical density from the 
kaon 2-point function to one-loop order and show that it 
is equal to that obtained from expanding the zero point energy.

Close to threshold, the grand potential may be expanded in 
terms of the condensate as follows 
\be
\frac{\Omega_K}{\vol} 
= \frac{\Omega_K^{(\theta=0)}}{\vol} -  
\thalf D^{-1}(\mu) f^2\theta^2  + {\cal O}(\theta^4)
\label{threshold}
\ee
where $D$ is the zero-momentum kaon propagator, relevant 
for {\it s}-wave condensation, and $\mu$ is the 
lepton chemical potential prior to kaon condensation. 
The threshold for kaon condensation is therefore given 
by  $D^{-1}(\mu)=0$, {\it i.e.} the density at which
the energy of negatively charged kaons
equals the lepton chemical potential.

{}From the chiral Lagrangian, Eq. (\ref{lkaon}), we may 
read off the interactions. 
The 2-point interactions are
\be
{\cal L}_{U2} = - (m_K^2+c) K^+K^-  \,\,\,,\,\,\,
{\cal L}_{V2} = -i b K^+\stackrel{\leftrightarrow}{\del^0}K^- \,\,.
\label{tps}
\ee
Here we treat the standard mass term as an interaction and
take the kaon propagator to be massless.  
The 4-point interactions are \cite{lbmr}  
\be
{\cal L}_{H4} &=& \frac{1}{6f^2} 
( K^+\stackrel{\leftrightarrow}{\del^\nu}K^- )^2 \,\,,\,\,
{\cal L}_{V4} = -\frac{ib}{6f^2} 
( K^+\stackrel{\leftrightarrow}{\del^0}K^- )K^+K^- \nonumber\\
{\cal L}_{U4} &=& \frac{(m_K^2+c)}{6f^2} ( K^+K^- )^2 \,\,.
\ee

We compute the kaon 2-point function with external four-momentum 
$p=(\omega,{\bf 0})$.  
The interactions (\ref{tps}) contribute at tree-level. 
To one-loop order, we have the diagrams of Fig. 3.
The contributions of Figs. 3(a), 3(b) and 3(c) are
$-\omega^2/(3f^2)$, $-2b\omega/(3f^2)$
and $2(m_K^2+c)/(3f^2)$, respectively.  
In the last two diagrams, the internal lines contain the 
interactions given in Eq. (\ref{tps}), which
are denoted by $m_K$ and $b$, respectively. 
Fig. 3(d) contributes $-(m_K^2+c)/(3f^2)$, and
Fig. 3(e) contributes $2b\omega/f^2$.
In addition, we must include all possible insertions of 
the interactions (\ref{tps}) on internal lines. 
Including the counterterm contribution, the result is
\be
D^{-1}(\omega) &=& 
\omega^2 + 2 b \omega - m_K^2 -c
+ (\omega^2 - 4\omega b-m_K^2-c) \delta  \quad{\rm where}\\
\delta &=&
- \frac{1}{6f^2} \left( \int \frac{d^3k}{(2\pi)^3}
\frac{1}
{\sqrt{{\bf k}^2+\xi^2}}
- 2 (d_2+f_2) + 2 \xi^2 (d_0+f_0) \right)
\label{Dinverse}
\ee
Eq. (\ref{Dinverse})
agrees with the expansion of $V_{zp}+V_{ct}$ given in Eq. (\ref{uc}),
as required, and in vacuum we obtain the free inverse propagator
$D^{-1}(\omega)=(\omega^2-m_K^2)$.

Note that Figs. 3(a), (c), and the contribution 
to (d) proportional to $m_K^2$, do not involve nucleon
fields and are mass and wavefunction renormalization
diagrams in the vacuum.  
These diagrams are ${\cal O}(q^4)$.
If we promote the nucleon fields to dynamical 
ones, we generate nucleon legs on Figs. 3(b) and (e),
corresponding to Figs. 1(g) and 1(f) respectively,  
and find that they are  ${\cal O}(q^3)$.
These diagrams correspond to Figs. 1(a) and 1(b)
of Ref.\cite{lbmr} respectively; since only nucleon and kaon
fields are used in the present calculation, the remaining diagrams
of Ref.\cite{lbmr} do not enter.

%
%

\begin{figure}
\leavevmode
\epsfbox[ 25 -50 502 488]{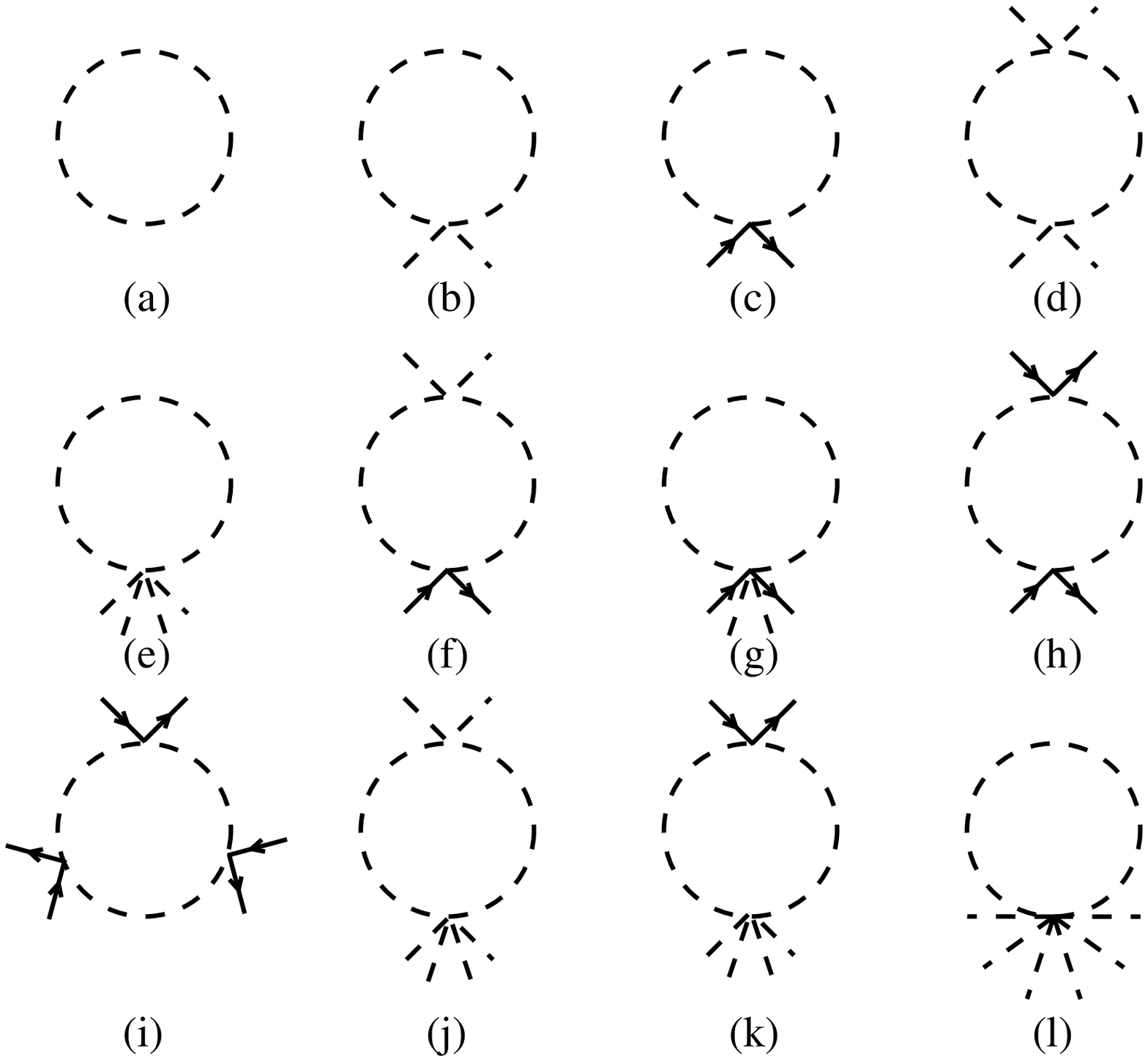}
\caption{Some contributions to the zero-point energy (effective potential 
to one-loop
order).  The dashed (solid) lines denote kaons (nucleons)
and the external legs are mean fields.
There are 5 additional diagrams with 6 external legs.}
\end{figure}

\begin{figure}
\leavevmode
\epsfysize=18cm
\epsfbox{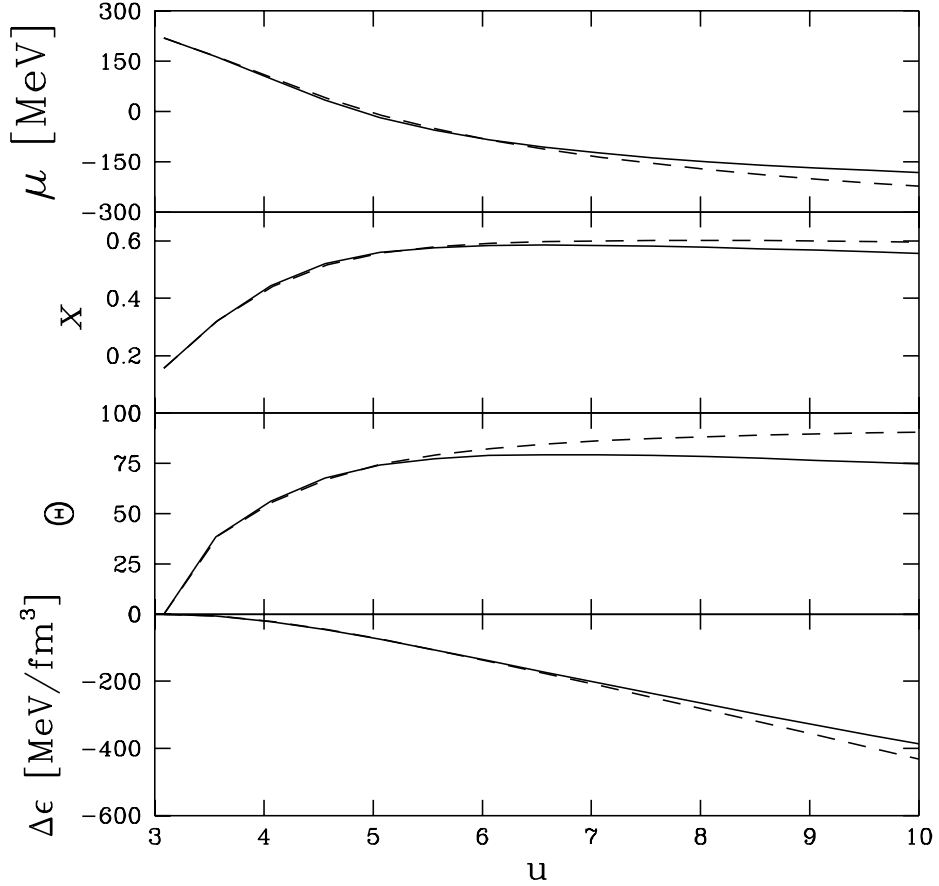}
\caption{Ground state neutron star properties for matter containing
a kaon condensate with $a_3m_s=-222$ MeV and $c=0$ in the loop
contribution. The dashed lines give the tree-level results and 
the solid lines show the effect of also including
the loop contribution.
Plotted are the charge chemical potential $\mu$, the proton fraction $x$, 
the condensate amplitude $\theta$, and the energy gained in forming a 
condensate 
$\Delta \epsilon$, as a function of  $u=n/n_0$, the density in units 
of equilibrium nuclear matter density.}
\end{figure}

\begin{figure}
\leavevmode
\epsfbox[ 50 -50 470 266]{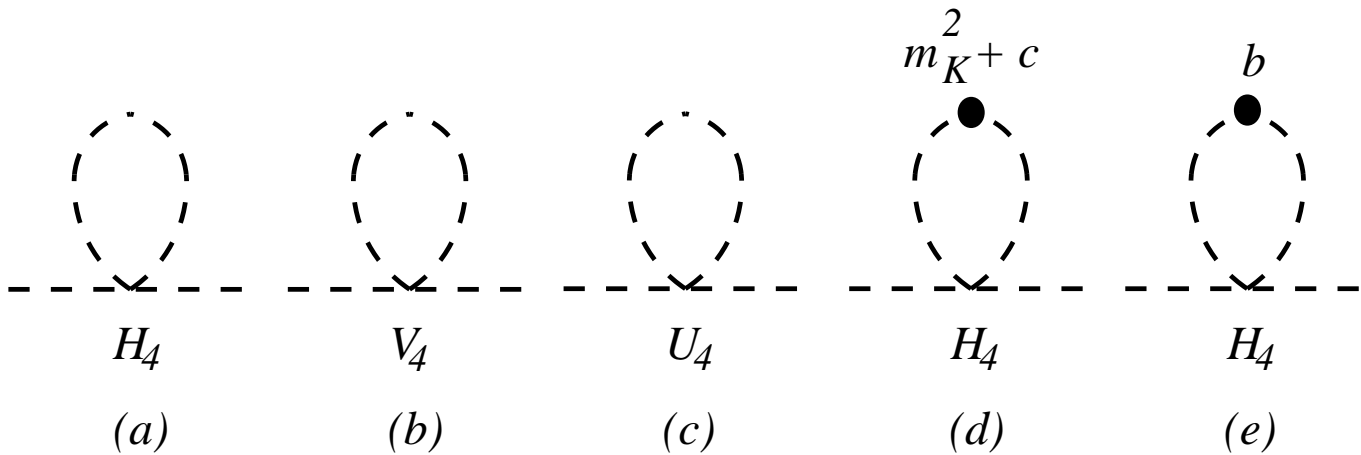}
\caption{Contributions to the kaon propagator to one-loop order. The dashed 
line denotes a free kaon propagator. See Appendix C for further 
notation.}
\end{figure}

\narrowtext 

\begin{table}
\caption{
Values of the critical density for kaon condensation, $u_{cr}=n_{cr}/n_0$, in
units of equilibrium nuclear matter density, for various choices of 
the parameter $a_3m_s$.  Superscript $t$ denotes the
tree level contribution, and $l$ denotes the loop
contribution. The scalar kaon-nucleon
interaction $c$ is defined in Eq. (5), while $c_{n_s}$ is
the corresponding quantity with the density $n$ replaced
by the scalar density $n_s$ given in Eq.(50)
} 
\begin{tabular}{lrrcrr}
 & & &$c=0$ in $l$ &\multicolumn{2}{c}{$c \rightarrow c_{n_s}$}\\
$a_3m_s$ &  $u_{cr}^{t}$  & $u_{cr}^{t+l}$ & $u_{cr}^{t+l}$ & 
$u_{cr}^{t}$ &  $u_{cr}^{t+l}$ \\
\tableline
$-$134 {\rm MeV} & 4.181 & 4.175 & 4.179 & 5.069 & 5.069\\
$-$222 {\rm MeV} & 3.082 & 3.064 & 3.081 & 4.231 & 4.224\\
$-$310 {\rm MeV} & 2.417 & 2.397 & 2.417 & 3.473 & 3.454\\
\end{tabular}
\end{table}

\begin{table}
\caption{
Properties of normal, $n,p,e^-,\mu^-$, neutron
star matter in the presence of kaon zero-point fluctuations as a function of 
density expressed in units of equilibrium nuclear matter density, $u=n/n_0$.
Here $a_3m_s=-222$ MeV.
Tabulated are the proton fraction $x$, 
the charge chemical potential $\mu$ (in MeV), 
the zero-point contribution to the energy
density $\epsilon_{zp}$ (${\rm MeV}\,{\rm fm}^{-3}$), 
and the total energy density $\epsilon$
(${\rm MeV}\,{\rm fm}^{-3}$).
}
\begin{tabular}{lrrrr}
  $u$  & $x$  &  $\mu$ &  $\epsilon_{zp}$  & $\epsilon$ \\
\tableline
 0.50 &0.015 &   64.8 &  $-$0.01 &  1.31 \\
 1.00 &0.039 &  110.9 &  $-$0.05 &  4.56 \\
 1.50 &0.074 &  145.2 &  $-$0.16 &  9.28 \\
 2.00 &0.106 &  172.7 &  $-$0.37 & 15.10 \\
 2.50 &0.132 &  195.9 &  $-$0.70 & 21.80 \\
 3.00 &0.154 &  216.2 &  $-$1.17 & 29.23 \\
 3.06 &0.156 &  218.6 &  $-$1.24 & 30.23 \\
%
\end{tabular}
\end{table}

\mediumtext 

\begin{table}
\caption{Ground state neutron star matter properties for 
normal $n,p,e,\mu$ matter, and
matter containing a kaon condensate to tree level
and one-loop level. Here $a_3m_s=-134$ MeV. 
Tabulated are the density in units of equilibrium nuclear matter density, $u$,
the amplitude of the condensate $\theta$ (in degrees), 
the charge chemical potential $\mu$ (MeV), 
the proton fraction $x$, 
the kaon fraction $x_K$, 
and the total energy density $\epsilon$
(${\rm MeV}\,{\rm fm}^{-3}$).
The zero-point contribution to
the energy density is $\epsilon_{zp}$ 
(${\rm MeV}\,{\rm fm}^{-3}$).
}
\begin{tabular}{ccccccccccccc}
    & $n,p,e,\mu$ & \multicolumn{5}{c}{$n,p,e,\mu,K$(tree)} & 
\multicolumn{6}{c}{$n,p,e,\mu,K$(tree+loop)} \\  
$u$ & $\epsilon$&   $\theta$ & $\mu$  & $x$ & $x_K$ & $\epsilon$ & 
                      $\theta$ & $\mu$  & $x$ & $x_K$ & $\epsilon_{zp}$ & 
							$\epsilon$      \\   
\tableline
 4.18& 51.98&  0.00& 256.3&0.194&0.000&  51.98&  
0.00& 256.1&0.194&0.000& $-$0.10&  51.88\\
 4.20& 52.36&  5.67& 255.3&0.198&0.007&  52.36&  
6.56& 254.7&0.199&0.010& $-$0.11&  52.25\\
 4.70& 62.60& 29.55& 226.9&0.281&0.166&  59.87& 
29.76& 225.4&0.283&0.170& $-$0.23&  59.64\\
 5.20& 73.40& 41.02& 197.7&0.346&0.280&  63.31& 
41.01& 195.8&0.348&0.284& $-$0.35&  62.95\\
 5.70& 84.74& 49.16& 168.8&0.395&0.361&  63.33& 
48.99& 166.8&0.396&0.363& $-$0.41&  62.91\\
 6.20& 96.57& 55.18& 141.3&0.431&0.415&  60.67& 
54.90& 139.5&0.431&0.416& $-$0.40&  60.27\\
 6.70&108.87& 59.63& 115.9&0.458&0.451&  56.03& 
59.31& 114.5&0.458&0.451& $-$0.32&  55.70\\
 7.20&121.61& 62.91&  92.9&0.477&0.474&  49.97& 
62.61&  92.1&0.477&0.474& $-$0.21&  49.75\\
 7.70&134.78& 65.42&  72.4&0.492&0.490&  42.90& 
65.20&  72.0&0.491&0.490& $-$0.11&  42.79\\
 8.20&148.35& 67.38&  54.2&0.502&0.502&  35.11& 
67.25&  54.1&0.502&0.502& $-$0.04&  35.08\\
 8.70&162.30& 68.92&  38.0&0.511&0.510&  26.80& 
68.87&  38.0&0.510&0.510& $-$0.01&  26.80\\
\end{tabular}
\end{table}

\mediumtext 

\begin{table}
\caption{Ground state neutron star matter properties for 
normal $n,p,e,\mu$ matter, and
matter containing a kaon condensate to tree level
and one-loop level.
Here $a_3m_s=-222$ MeV (See Table III for notation).
}
\begin{tabular}{ccccccccccccc}
    & $n,p,e,\mu$ & \multicolumn{5}{c}{$n,p,e,\mu,K$(tree)} & 
\multicolumn{6}{c}{$n,p,e,\mu,K$(tree+loop)} \\
$u$ & $\epsilon$&   $\theta$ & $\mu$  & $x$ & $x_K$ & $\epsilon$ & 
                      $\theta$ & $\mu$  & $x$ & $x_K$ & $\epsilon_{zp}$ & 
							$\epsilon$      \\   
\tableline
 3.06& 31.47&      &      &     &     &       &  
0.00& 218.6&0.156&0.000& $-$1.24&  30.23\\
 3.08& 31.77&  0.00& 219.2&0.157&0.000&  31.77&  
7.20& 216.7&0.163&0.012& $-$1.27&  30.50\\
 3.26& 34.82& 22.86& 200.0&0.222&0.113&  34.10& 
24.31& 196.3&0.230&0.128& $-$1.67&  32.44\\
 3.46& 38.35& 33.81& 177.1&0.287&0.220&  35.13& 
34.88& 172.2&0.296&0.235& $-$2.17&  32.97\\
 3.66& 41.99& 42.29& 152.8&0.345&0.308&  34.52& 
43.03& 147.0&0.353&0.321& $-$2.70&  31.82\\
 3.86& 45.75& 49.41& 127.6&0.395&0.378&  32.36& 
49.71& 121.3&0.401&0.387& $-$3.18&  29.17\\
 4.06& 49.60& 55.40& 102.0&0.438&0.430&  28.81& 
55.15&  96.0&0.441&0.435& $-$3.56&  25.22\\
 4.26& 53.56& 60.50&  76.8&0.473&0.470&  24.06& 
59.68&  71.7&0.473&0.470& $-$3.80&  20.21\\
 4.46& 57.61& 64.91&  52.6&0.501&0.500&  18.30& 
63.52&  48.9&0.498&0.498& $-$3.90&  14.35\\
 4.66& 61.76& 68.66&  29.9&0.524&0.524&  11.69& 
66.77&  27.8&0.519&0.519& $-$3.85&   7.80\\
 4.86& 66.00& 71.80&   9.0&0.542&0.542&   4.41& 
69.53&   8.3&0.535&0.535& $-$3.69&   0.71\\
 5.06& 70.32& 74.43& $-$10.3&0.556&0.556&  $-$3.40& 
71.89&  $-$9.7&0.548&0.548&
 $-$3.44&  $-$6.82\\
\end{tabular}
\end{table}

\mediumtext 

\begin{table}
\caption{Ground state neutron star matter properties for 
normal $n,p,e,\mu$ matter, and
matter containing a kaon condensate to tree level
and one-loop level.
Here $a_3m_s=-222$ MeV and
the scalar density of Eq. (50) 
is employed in the scalar 
kaon-nucleon interaction $c_{n_s}$.
(See Table III for notation.)}
\begin{tabular}{ccccccccccccc}
    & $n,p,e,\mu$ & \multicolumn{5}{c}{$n,p,e,\mu,K$(tree)} & 
\multicolumn{6}{c}{$n,p,e,\mu,K$(tree+loop)} \\
$u$ & $\epsilon$&   $\theta$ & $\mu$  & $x$ & $x_K$ & $\epsilon$ & 
                      $\theta$ & $\mu$  & $x$ & $x_K$ & $\epsilon_{zp}$ & 
							$\epsilon$      \\   
\tableline
 4.22& 52.84&      &      &     &     &       &  
0.00& 257.6&0.195&0.000& 
$-$0.09&  52.75\\      
 4.23& 52.96&  0.00& 257.8&0.196&0.000&  52.96&  
2.56& 257.4&0.196&0.001& $-$0.09&  52.88\\
 4.30& 54.36&  8.93& 255.5&0.205&0.017&  54.28&  
9.57& 255.0&0.206&0.020& $-$0.09&  54.24\\
 4.80& 64.72& 25.67& 239.0&0.263&0.128&  62.86& 
25.86& 238.1&0.264&0.131& $-$0.11&  62.79\\
 5.30& 75.63& 34.53& 223.2&0.308&0.210&  69.54& 
34.51& 222.2&0.309&0.213& $-$0.13&  69.44\\
 5.80& 87.07& 40.88& 208.2&0.342&0.272&  74.69& 
40.73& 207.3&0.343&0.274& $-$0.12&  74.59\\
 6.30& 98.99& 45.75& 194.3&0.369&0.318&  78.63& 
45.51& 193.5&0.370&0.319& $-$0.10&  78.55\\
 6.80&111.38& 49.58& 181.3&0.390&0.353&  81.64& 
49.29& 180.8&0.391&0.354& $-$0.08&  81.58\\
 7.30&124.21& 52.64& 169.5&0.407&0.380&  83.94& 
52.34& 169.2&0.407&0.380& $-$0.06&  83.89\\
 7.80&137.46& 55.11& 158.8&0.421&0.400&  85.69& 
54.82& 158.6&0.421&0.401& $-$0.05&  85.65\\
 8.30&151.11& 57.12& 149.1&0.431&0.416&  87.03& 
56.85& 149.0&0.431&0.417& $-$0.04&  86.99\\
 8.80&165.13& 58.77& 140.3&0.440&0.429&  88.05& 
58.52& 140.1&0.440&0.429& $-$0.02&  88.03\\
 9.30&179.53& 60.15& 132.4&0.447&0.439&  88.84& 
59.89& 131.8&0.447&0.439&  0.04&  88.88\\
\end{tabular}
\end{table}


\begin{references}


\bibitem[*]{byline} Permanent address.

\bibitem{bat} E. Friedman, A. Gal and C. J. Batty,
Nucl. Phys. {\bf A579}, 578 (1994).

\bibitem{kapnel}
D.~B. Kaplan and A.~E. Nelson,
Phys. Lett. {\bf B175}, 57 (1986) ; {\bf B179}, 409 (E) (1986). 

\bibitem{blrt}
G.~E. Brown, C.-H.~Lee, M.~Rho and V.~Thorsson, Nucl. Phys. {\bf A567}, 
937 (1994). 

\bibitem{lbmr} 
C.-H. Lee, G. E. Brown, D.-P. Min and M. Rho,
Nucl. Phys. {\bf A585}, 401 (1995).

\bibitem{ppt}
V. R. Pandharipande, C. J. Pethick and V. Thorsson,
Phys. Rev. Lett. {\bf 75}, 4567 (1995);
T. Waas, M. Rho and W. Weise, Preprint {\tt nucl-th/9610031},
Submitted to Nucl. Phys.       

\bibitem{kyoto}
T. Muto and T. Tatsumi, Phys. Lett. {\bf B283}, 165 (1992);
H. Fujii, T. Maruyama, T. Muto, and T. Tatsumi
Nucl. Phys. {\bf A597}, 645 (1996).

\bibitem{waas}
T. Waas, N. Kaiser, W. Weise, Phys. Lett. {\bf B365}, 12 (1996);
{\bf B379}, 34 (1996).

\bibitem{likoli}
G. Q. Li, C. M. Ko and B.-A. Li, Phys. Rev. Lett. {\bf 74}, 235 (1995).

\bibitem{subt}
G. Q. Li, C. M. Ko and X. S. Fang, Phys. Lett. {\bf B329}, 149 (1994);
A. Schr\"{o}ter {\it et al.}, Z. Phys. {\bf A350}, 101 (1994);
J. Schaffner,  J. Bondorf and I.N. Mishustin, 
NBI Preprint 96-41, {\tt nucl-th/9607058}.

\bibitem{phi}
T. Tatsumi, H. Shin, T. Maruyama, and H. Fujii,
KUNS-1383, Aust. J. Phys., to be published.

\bibitem{hyp} P.J. Ellis, R. Knorren and M. Prakash,
Phys. Lett. {\bf B349}, 11 (1995); R. Knorren, M.
Prakash and P.J. Ellis, Phys. Rev. {\bf C52}, 3470 (1995);
J. Schaffner and I.N. Mishustin, Phys. Rev. {\bf C53}, 1416 (1996);
J. Schaffner,  J. Bondorf and I.N. Mishustin, Strangeness '96
(Budapest, Hungary, 1996) Heavy Ion Physics, to be pub., 
{\tt nucl-th/9607019}.

\bibitem{bkrt}
G.~E. Brown, K.~Kubodera, M.~Rho and V.~Thorsson, Phys. Lett. {\bf B291}, 
355 (1992). 

\bibitem{tpl}
V. Thorsson, M. Prakash and J. M. Lattimer, Nucl. Phys. {\bf A572}, 693 (1994);
{\bf A574}, 851 (1994). 

\bibitem{Taylor} S. E. Thorsett, Z. Arzoumanian, M. M. McKinnon 
and J. H. Taylor, Astrophys. J. {\bf 405}, L29 (1994).

\bibitem{keiljan} W. Keil and H.-Th. Janka,
Astron. and Astrophys. {\bf 296}, 145 (1994).

\bibitem{pcl} M. Prakash, J. Cooke and J.M. Lattimer,
Phys. Rev. {\bf D52}, 661 (1995).

\bibitem{subside} N.K. Glendenning,
Astrophys. J. {\bf448}, 797 (1995).

\bibitem{newborn}
M. Prakash, I. Bombaci, M. Prakash, P. J. Ellis, J. M. Lattimer
and R. Knorren, {\tt nucl-th/9603042}, Phys. Rep. (1996) in press;
P. J. Ellis, J. M. Lattimer and M. Prakash, 
Comments on Nucl. and Part. Phys. {\bf22}, 63 (1996);
M. Prakash, S. Reddy, J.M. Lattimer and P.J. Ellis, Strangeness '96
(Budapest, Hungary, 1996) Heavy Ion Physics, to be pub., 
{\tt nucl-th/9607031}.

\bibitem{bb}
G. E. Brown and H. A. Bethe, Astroph. J. {\bf 423}, 659 (1994).

\bibitem{burr} A. Burrows, Astrophys. J. {\bf334}, 891 (1988).

\bibitem{bst}
T. W. Baumgarte, S. L. Shapiro, and S. Teukolsky, Astroph. J. {\bf 443},
717 (1995); {\it ibid} {\bf 458}, 680 (1996).  

\bibitem{jm} E. Jenkins and A. V. Manohar, 
Phys. Lett. {\bf B255}, 558 (1991).

\btem{dong} S-J. Dong and K-F. Liu,
Nucl. Phys. {\bf B} (Proc. Suppl.) {\bf 42}, 322 (1995).


\btem{kapusta} J. I. Kapusta,
{\em Finite Temperature Field Theory} (Cambridge University Press, 1985).

\btem{bbd} K. M. Benson, J. Bernstein and S. Dodelson, 
Phys. Rev. {\bf D44}, 2480 (1991).

\btem{pokorski}
S. Pokorski, {\em Gauge Field Theories} (Cambridge University Press, 1987).

\btem{wein} S. Weinberg, Phys. Lett. {\bf B251}, 288 (1990);
Nucl. Phys. {\bf B363}, 2 (1991).

\bibitem{tw}
V. Thorsson and A. Wirzba, Nucl. Phys. {\bf A589}, 633 (1995).

\btem{as} M. Abramowitz and I. A. Stegun, {\it Handbook of Mathematical 
Functions} (Dover, NY, 1965).

\btem{pal} M. Prakash, T. L. Ainsworth and J. M. Lattimer,
Phys. Rev. Lett. {\bf 61}, 2518 (1988).

\bibitem{nsvsnb}
J. Schaffner, A. Gal, I. N. Mishustin, H. Stocker, and W. Greiner, 
Phys. Lett. {\bf B334}, 268 (1994); T. Maruyama, H. Fujii, 
T. Muto, and T. Tatsumi, Phys. Lett. {\bf B337}, 19 (1994).

\bibitem{sw}
B. D. Serot and J. D. Walecka, Advances in Nucl. Phys. {\bf 16}, ed. J. W.
Negele and E. Vogt (Plenum, NY, 1986).

\bibitem{wwu} 
E. J. Weinberg and A. Wu, Phys. Rev. {\bf D36}, 2474 (1987).

\bibitem{mm}
S. G. Matinyan and B. M\"uller, Preprint 
DUKE-TH-96-132, Contribution to 'Foundations in Physics' 
commemorating L. C. Biedenharn, Submitted to Found. Phys, {\tt hep-th/9610233}.

\btem{gl}
J. Gasser and H. Leutwyler, Nucl. Phys. {\bf B250}, 465 (1985).

\btem{zwm}
I. Zahed, A. Wirzba and U.-G. Meissner, Phys. Rev. {\bf D33}, 830 (1986)

\end{references}
\end{document}